\documentclass[12pt,preprint]{aastex}
\usepackage{lscape}
\usepackage{epsfig}
\usepackage{ulem}
\usepackage{soul}
\newcommand \kms {\ifmmode{\rm km\thinspace s^{-1}}\else km\thinspace s$^{-1}$\fi}

\begin{document}

\title{A Photometric Survey for Variables and Transits in the Field of Praesepe with KELT}
\author 
{Joshua Pepper\altaffilmark{1,2}, K.~Z. Stanek\altaffilmark{1}, Richard W. Pogge\altaffilmark{1}, David W. Latham\altaffilmark{3}, D.~L. DePoy\altaffilmark{1}, Robert Siverd\altaffilmark{1}, Shawn Poindexter\altaffilmark{1}, Gregory R. Sivakoff\altaffilmark{1}}
\altaffiltext{1}{The Ohio State University Department of Astronomy, 4055 McPherson Lab, 140 West 18th Ave., Columbus, OH 43210}
\altaffiltext{2}{Current Address: Physics \& Astronomy Department, Vanderbilt University, 6907E Stevenson Center, Nashville, TN 37235; joshua.pepper@vanderbilt.edu}
\altaffiltext{3}{Harvard-Smithsonian Center for Astrophysics, 60 Garden Street, Cambridge, MA 02138}

\begin{abstract}
The Kilodegree Extremely Little Telescope (KELT) project is a small aperture, wide-angle search 
for planetary transits of solar-type stars.  In this paper, we present the results of a 
commissioning campaign with the KELT telescope to observe the open cluster Praesepe for 34 
nights in early 2005.  Lightcurves were obtained for 69,337 stars, out of which we identify 
58 long period variables and 152 periodic variables.  Sixteen of these are 
previously known as variable, yielding 194 newly discovered variable stars for which we provide 
properties and lightcurves.  We also searched for planetary-like transits, finding four transit 
candidates.  Follow-up observations indicate that two of the candidates are astrophysical 
false positives, with two candidates remaining as potential planetary transits.
\end{abstract}


\section{Introduction} \label{sec:intro}

The field of planet searches has grown tremendously in the past several years.  One of 
the techniques for planet detection that has had recent successes is the search for 
planets transiting their host stars.  Transits of bright stars have great 
scientific potential, giving clues to the internal structure of 
planets \citep{g05}, their atmospheric composition \citep{char02}, spin-orbit 
alignment \citep{gw07}, and the presence of rings or 
moons \citep{bf04} -- see \citet{char07} for a comprehensive review.  

To date 18 transiting planets are known.  Five of them were discovered first through 
radial-velocity searches and were then found to be transiting, while the rest were 
discovered by photometric transit surveys.   Of those found using the transit method, five 
were found by the Optical Gravitational Lensing Experiment (OGLE) 
survey \citep{ud02a,ud02b,ud02c,ud03,ud04} but have relatively faint ($V > 16$) host 
stars.  The eight remaining planets orbit relatively bright ($V \leq 12 $) stars and were 
discovered by small telescopes with wide fields of view 
\citep{alonso04,bakos07,burke07,cam07,mc06,od06,od07}.

The Kilodegree Extremely Little Telescope (KELT) project is a wide-field, small-aperture 
survey for planetary transits of stars with $8 < V < 10$ mag.  It is similar to other 
wide-field transit surveys such as SuperWASP \citep{p06}, XO \citep{mc05}, HAT \citep{bakos04}, 
and TrES \citep{alonso04}.  The justification for the 
parameters of our survey strategy is described in \citet{pep03}, and the instrumentation, 
performance, and observing strategy are described in \citet{pep07}.  

In this paper, we report the results of our commissioning observations.  These data were 
used to establish our operational procedures and to build and test the software pipeline.  The 
purpose was not to discover transits with these observations, but rather to use the data to 
build the analytical tools for use with comprehensive surveys with KELT.  The choice to 
observe the cluster Praesepe was made based on convenience, as an accessible target at the 
time of commissioning, and the possibility for scientific potential from studying the 
variable stars in the cluster.  The details of the data set analyzed here are not 
identical to the main KELT survey data, but we can use the observations to test our 
ability to obtain transit-quality photometry, defined as lightcurves with low 
noise, both random and systematic, able to discern astrophysical signals with 
the timescales and depth of typical planetary transits. 

We first briefly review the KELT 
instrument (\S \ref{sec:instr}) and describe the Praesepe 
observations (\S \ref{sec:obs}).  We then discuss the data reduction 
process, (\S \ref{sec:datar}), with special attention to the problems that affected 
the data quality (\S \ref{sec:focus}), and assess the photometric precision of the 
data set (\S \ref{sec:rms}).  We explain how we search for variable 
stars (\S \ref{sec:vars}) and transit candidates (\S \ref{sec:trans}).  We then list 
the properties of all variable stars detected and display the lightcurves of the 
periodic variables (\S \ref{sec:vcat}), and describe the 
final set of transit candidates and the observations to confirm their 
nature (\S \ref{sec:tcan}).  We conclude by reviewing the usefulness of this data 
set and the implications for the full KELT survey (\S \ref{sec:disc}).

\section{Instrumentation} \label{sec:instr}

Here we provide a summary of the full instrumental specifications of KELT that are described 
in \citet{pep07}.  The KELT telescope uses an Apogee Instruments 
AP16E thermoelectrically cooled CCD camera.  This camera uses a Kodak KAF-16801E front-side
illuminated CCD with $4096 \times 4096$ $9\mu$m pixels (36.88 $\times$ 36.88\,mm detector 
area).  It has a gain of 3.6\,electrons/ADU, readout noise of $\sim$15\,e$^-$, and 
saturates at 16383 ADU ($\sim$59,000 e$^-$), with very low dark current.  The camera is mounted 
on a Paramount ME Robotic Telescope Mount manufactured by Software 
Bisque.  The Paramount is a research-grade German Equatorial 
Mount designed specifically for robotic operation with integrated telescope and camera control.
For our observations of Praesepe, we use a Mamiya 645 200\,mm f/2.8 APO manual-focus 
telephoto lens with a 71\,mm aperture.  This provides a roughly 9\farcs5pix$^{-1}$ image 
scale and effective 10\fdg8$\times$10\fdg8 field of view.  A Kodak Wratten \#8 red-pass filter 
with a 50\% transmission point at $\sim$490\,nm, is mounted in front of the KELT lens.  The 
effective wavelength of the combined Filter+CCD response function (excluding atmospheric effects)
is 691nm, with an effective width of 318nm.  This results in an effective bandpass that is 
equivalent to a very broad R-band filter.

The KELT telescope is currently operated at the Irvin M. Winer Memorial Mobile 
Observatory\footnote{http://www.winer.org} near Sonoita, Arizona.  The site is located 
at N 31\degr39\arcmin53\arcsec, W 110\degr36\arcmin03\arcsec, approximately 50 miles 
southeast of Tucson at an elevation of 1515\,meters (4970\,feet).

\section{Observations} \label{sec:obs}

The commissioning campaign targeted a field centered on Praesepe (also called M44 and NGC 2632), 
an open cluster at a distance of 180 parsecs with an age of $600$ Myr, a metallicity 
slightly above solar, and little to no reddening \citep{an07}.  Several previous studies have 
examined the cluster population in efforts to determine the stellar luminosity function and to 
establish overall cluster membership \citep{js91,ad02}, in addition to probing the low-mass end 
of the stellar population \citep{ch05}.  The amount of interstellar extinction, $E(B-V)$, towards 
the center of Praesepe is 0.029 mag \citep{sfd98}, corresponding to $A_V = 0.09$ for $R_V = 3.1$.

Our observation were centered on a 10\fdg8$\times$10\fdg8 field located at 
J2000 $8^{\rm h}40^{\rm m}06^{\rm s}, +19^{\rm o}41^{'}06^{''}$, roughly centered on Praesepe.  The 
observing campaign was conducted every clear night from UTC 2005 February 13 until 2005 April 27, 
obtaining 5220 images during 34 out of the 74 nights of the run.
The observations consisted of 60-second observations repeated throughout 
the night as long as the cluster was above the horizon, resulting in 100 -- 200 images each 
night, with a 90-second cadence.  The telescope took images on all 
nights except those with heavy cloud cover or rain.  The 
quality of our observed nights ranged from completely clear to patchy cloud cover.  We rely 
on several steps of the data reduction process to eliminate images with excessive cloud cover, 
moonlight, or other problems that compromise photometric quality.

The pointing of the KELT telescope was not perfect, with the coordinates of the field center 
drifting slowly between images throughout the night.  The typical intranight drift 
was $\sim$25\arcmin\ ($\sim$160 pixels) in Declination and $\sim$9\arcmin\ ($\sim$60 pixels) 
in Right Ascension over the course of many hours.  The drift is small compared to the size of 
the field ($<5$\%), but it has two significant effects on our data.  First, the drift causes 
stars at the edges of the field to enter and exit the camera's field of view during the night, 
resulting in incomplete lightcurves for those stars.  However, since the image quality is poor 
at the extreme edges of the field, we simply 
eliminate stars along the field edges from our sample.  Secondly, the drift combines 
with our image reduction software to cause constant stars in certain parts of the 
field to spuriously appear as variable candidates.  See sections \S \ref{sec:imsub} 
and \S \ref{sec:cluster} for details of this effect.

\section{Data Reduction} \label{sec:datar}

For each of the 5220 images of Praesepe, we subtracted the combined dark for that 
night, and then divided by a flatfield.  Images showing large stellar image FWHM or very 
high (800 ADU) sky levels are eliminated as ``bad''.  These sky levels are mostly caused by 
clouds.  In all, 2083 poor-quality images were eliminated from further analysis.  The 
remaining 3137 images were analyzed using the ISIS image subtraction package 
from \citet{al98,alard00}.  We adopt the implementation of ISIS described 
by \citet{hart04}.  We describe that procedure below, and we note where our procedures differ 
from those of \citet{hart04}.

\subsection{Changing FWHM} \label{sec:focus}

After initial image processing with ISIS, we found that a large number of variable-star candidates 
had double-valued lightcurves, with the amplitude of many sinusoidal lightcurves 
being larger on some nights and smaller on others.  This variation in amplitude was correlated 
in time, and appeared to apply mostly to stars in a horizontal zone across the 
upper part of the chip.  We traced the origin of this effect to changing FWHM of the stellar 
images across our field from night to night.  The FWHM does not change significantly horizontally 
across the field, but it has significant structure in the Y-direction, with the FWHM being large 
($\sim2.8$ pixels) at the top and bottom of the chip, and decreasing linearly towards the middle 
of the chip in a V-shape, with a minimum FWHM of $\sim2.0$ pixels at about one third of the distance from the 
top of the chip.

Had the size and shape of this pattern remained constant throughout the observations, 
it would not have presented a major problem, since ISIS is able to work with a changing 
FWHM across the field.  However, the vertical-axis position of the FWHM minimum 
changed significantly between different nights, ranging from the middle of the chip to the top 
edge.  That is, the bottom of the V-shape of the FWHM distribution moved up and down the chip 
over different nights.  The position of the FWHM minimum correlated with the change in amplitude of the 
lightcurves.  

The reason for the double-valued lightcurves relates to the way ISIS works.  ISIS requires a 
reference image that has the smallest FWHM of all images in the set, since it convolves the 
reference image with the kernel of each of the individual images.  With the changing shape 
of the FWHM pattern in our data, no single image or set of images has a smaller FWHM across 
the entire field for all the nights.  Any given choice for a reference image will contain a 
horizontal region for which there are other nights where the stars have smaller FWHM 
values.  Regions with smaller PSFs than the reference image on certain nights require deconvolution 
of the PSFs (i.e. adjusting the stellar images from the reference frame to have smaller PSFs 
rather than larger).  This creates a problem since ISIS works to convolve an 
image, even with a varying degree across the field, but ISIS is not equipped to 
correctly {\it de}convolve an image.  

We have not been able to absolutely determine the origin of the time-varying FWHM pattern.  We 
believe it could be related to temperature changes from night to night, which affected 
the 200\,mm lens we used for these observations.  Since the problem was not detected until 
after we stopped using the lens, we were unable to test this hypothesis.  Instead, our objective 
is to mitigate the impact of this effect as much as possible and get the best measurements we can 
out of the data set.  

\begin{figure}[b]
\epsscale{1.0}
\plotone{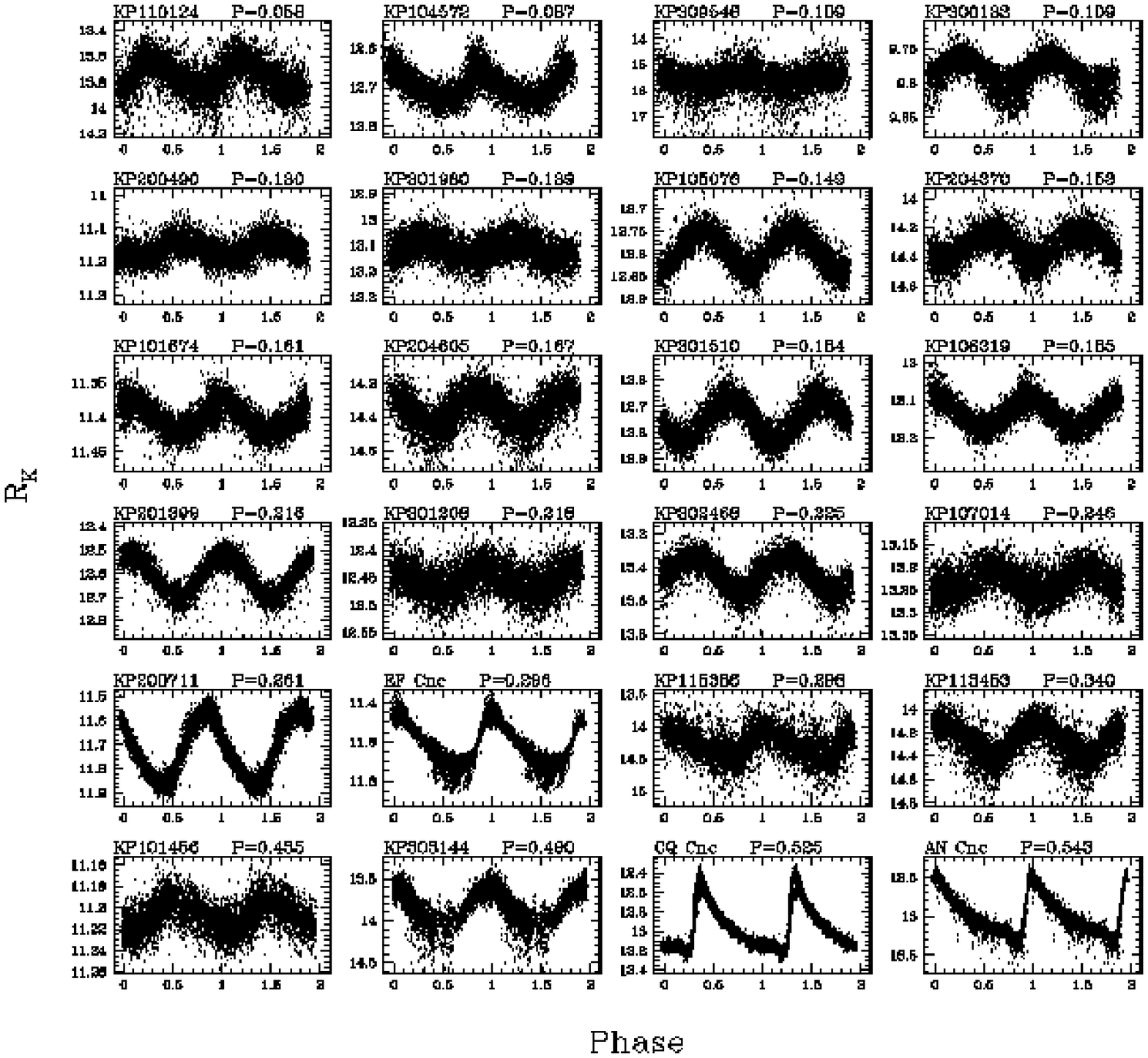}
\caption{Lightcurves of pulsating periodic variables identified by KELT, sorted by period.}
\label{fig:var.puls1}
\end{figure}

\begin{figure}[b]
\epsscale{1.0}
\plotone{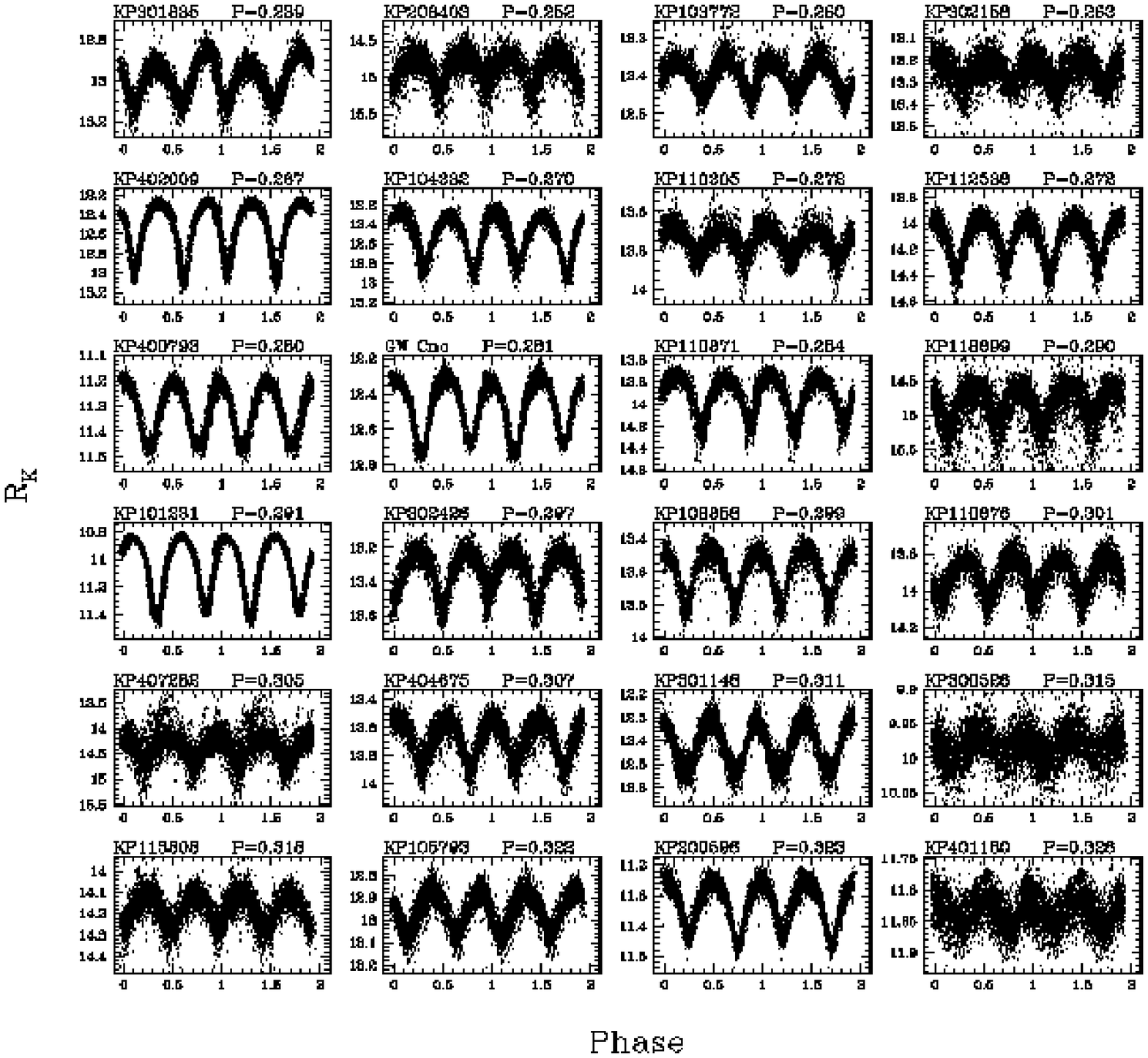}
\caption{Lightcurves of eclipsing periodic variables identified by KELT, sorted by period.}
\label{fig:var.eb1}
\end{figure}

The procedure we adopted is as follows.
We first register the images to the same coordinate system, then divide 
them into four horizontal sections.  For each image section we identify a different reference 
image that has the smallest FWHM out of all images in that section. We then treat the 
different sections as if they were four separate images, and run each section though the data reduction 
pipeline with their own reference images to obtain lightcurves (with an additional subdivision 
step described in \S \ref{sec:imsub}).  We also convolve all of the images (but not the reference 
image) with a Gaussian smoothing function (using a Gaussian with $\sigma=0.6$ pixels) to slightly broaden 
the PSFs, and thus ensure that the reference image has a smaller FWHM.  With this procedure, we are 
able to eliminate most of the 
effects of the time-varying FWHM.  We are not able to completely eliminate the effect, which 
can be seen in the slightly double-valued lightcurves of some variables (e.g., see plots 
of KP300133 and EF Cnc in Figure \ref{fig:var.puls1}, as well as excessive scatter in the 
lightcurves of KP113808 and KP118899 in Figure \ref{fig:var.eb1}).

\subsection{Image Subtraction} \label{sec:imsub}

We perform image subtraction on each section of the 3137 images with 
ISIS, as described in \S \ref{sec:focus}.  We use a feature of ISIS to further subdivide each 
section into a grid of 
subfields, each of which can take on different values for the parameters that are used 
to convolve the reference image with the kernel for image subtraction.  This step is 
particularly advantageous for large fields of view, in which cloud patterns can be smaller 
than the size of the field.  We subdivide each section into grids 
from $1 \times 5$ to $5 \times 5$, depending on the size of the section, and proceed with 
the image subtraction.

We use DAOPHOT \citep{stet87} to identify all of the objects in the field of the reference image of 
each section, yielding a list of 69,337 stars on all four sections.  Stars along the image edges have 
particularly bad lightcurves, and so we remove all stars within 50 pixels 
of the edge from further analysis, leaving 66,638 stars for which we generate lightcurves.  The ISIS 
photometry program calculates the flux from a star and its error in each image.  In some situations, 
such as when 
the star is located at the edge of the chip, or on the edge of one of the of subregions of the 
subtracted image, the reported flux and error values are not indicative of the true flux.  We 
clean such points from the data by removing data points where the reported flux or error is 
unphysically high or low.  We also remove the two highest and lowest flux measurements from each 
lightcurve.  Removing such a small number of points should not affect detection of 
variability or transits, but it does help remove spurious points and reduces the number of false 
positives when searching for variable sources.

\subsection{Astrometry and Matching to Known Sources} \label{sec:astrom}

We use the {\it Astrometrix}\footnote{\url{http://www.na.astro.it/$\sim$radovich/wifix.htm}} 
program to derive astrometric solutions for the reference 
image, using the Tycho-2 catalog \citep{hog00} to select reference stars.  Because of 
high-order distortions in the corners of the field, we first subdivide the full images 
into 25 subsections and find separate astrometric solutions  for each subimage.  The 
astrometry is good to within an arcsecond, or $\sim 0.1$ pixel.

We match our data set to two catalogs.  We first match our stars to the 2MASS 
catalog \citep{sk06}, using a search radius of 9.5 arcseconds or about one KELT pixel.  We 
find that 58,620 out of 66,638 of our KELT stars are in the 2MASS catalog, with 1,559 of them matching to 
more than one 2MASS source.

We also match our star catalog to known members of the Praesepe cluster.  We compiled a catalog 
from the WebDA website, identifying 832 likely cluster members.  After matching to the KELT 
data using a search radius of 9.5 arcseconds, we find matches to 333 Praesepe member 
stars.  However, many of the stars in the WebDA database are too faint for KELT to 
detect.  If we consider only 
the 210 WebDA sources with known $V$ magnitudes of $6.8 < V < 16.4$, we find matches to 
147 stars, although the brighter stars are mostly saturated and unusable in the KELT images.

\subsection{Photometric Calibration} \label{sec:calib}

Our goal for KELT is to obtain highly precise relative photometry, so we 
do not attempt to achieve extremely precise absolute photometry.  The KELT bandpass 
is an approximate wide $R$ band.  We define a KELT magnitude $R_K$ to which we calibrate 
our observations, which is within a few tenths of a magnitude of Johnson $R$ for the bulk of our 
stars.  However, because of our broad filter and wide field, we are susceptible 
to significant color terms when determining absolute photometry.  For stars with a 
known $(V-I)$ color we can determine the $V$ magnitude to within a tenth of a magnitude.  Since 
we do not know the $(V-I)$ colors of most of our stars, we quote all our observed 
magnitudes in $R_K$, which can be considered to be equivalent to Johnson $R$, modulo a 
color term which is typically 0.2 magnitudes, but can range from $-0.3$ magnitudes for very 
blue stars to 0.8 magnitudes for very red stars.  See \citet{pep07} for full details about 
the calibration process.

\subsection{Rescaling Errors} \label{sec:errs}

One feature of ISIS that has been noted by others is that the formal reported errors 
tend to be underestimated for brighter stars.  Since the errors on individual points 
are important in the variable selection process, we rescale the errors following the 
procedure of \citet{kal98} and \citet{hart04}.  We first compute the reduced $\chi^2$ for every star 
\begin{equation} \label{equ:chi}
\chi^2/N_{\rm dof} = \frac{1}{N-1} \sum_{k=1}^{N} \frac{(m_k - \mu)^2}{\sigma_k^2},
\end{equation}
where the sum is over $N$ observations, $m_k$ is the instrumental magnitude with error $\sigma_k$,
and $\mu$ is the weighted-average instrumental magnitude.
We then plot $\chi^2/N_{\rm dof}$ 
versus magnitude and fit a curve to the bottom edge of the heaviest concentration 
of points.  We then multiply the formal errors by the square root of the function of
that curve, so that the least variable lightcurves in our data set 
have $\chi^2/N_{\rm dof}$ close to 1 for all magnitudes.

\subsection{Photometric Precision} \label{sec:rms}

\begin{figure}[b]
\epsscale{1.0}
\plotone{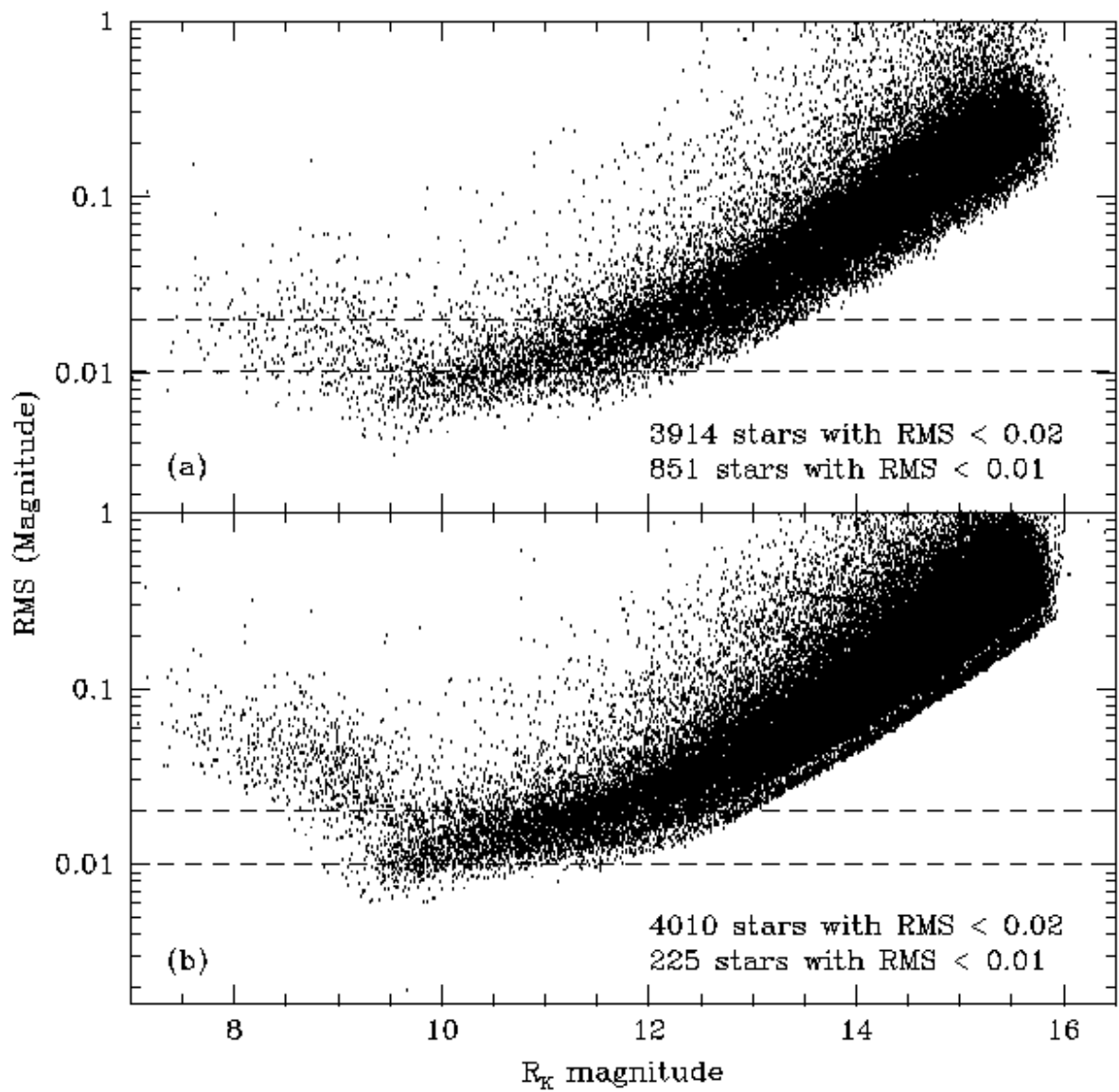}
\caption{(a) RMS vs. $R_K$ magnitude for the 66,638 KELT stars, calculated from 105 
observations over the course of one night. (b) RMS plot for the same stars 
using all 3,137 observations over the full 34 nights.  Structure in the 
plots results from the separate reductions of the four sections of the field to deal 
with the changes in the FWHM pattern.}
\label{fig:rms}
\end{figure}

A common method for describing the photometric precision of transit searches 
is to plot the magnitude root-mean-squared (RMS) values of all the lightcurves as a 
function of magnitude.  Because of the FWHM changes described in \S \ref{sec:focus}, the 
long term photometric precision has been degraded.  However, since the intranight 
FWHM pattern appears to be stable, we plot the overall RMS and the
RMS for one of the nights for all 66,638 stars in Figure \ref{fig:rms}.  Panel (a) of Figure \ref{fig:rms} 
shows the RMS plot for a single night of data, while panel (b) shows the RMS plot 
for the full 34 nights of data.  The dashed horizontal lines show the 2\% 
and 1\% RMS limits, which generally define the required sensitivity for detecting 
Hot Jupiter transits.  Two features stand out in these plots.  The first is that 
there are significantly fewer stars with RMS $< 1$\% in plot (b) than plot (a), although 
there are about the same number of stars in each plot with RMS $< 2\%$.  This behavior 
shows that the inter-night systematics, of which we believe the FWHM changes to be 
the most significant, become most important at the sub-1\% level.  The second feature to 
note is the greater RMS of stars brighter than $R_K \approx 9$.  The lightcurves of the 
brightest stars in our sample are dominated by systematic saturation and/or nonlinearity 
effects that are present during a single night but are more severe over the entire run. 

\section{Variable Selection} \label{sec:vars}

Any transit search will yield a data set suitable for detecting variable stars that are 
unrelated to transiting planets.  We 
implement several cuts to select promising variable star candidates.  We first employ 
the Stetson $J$ statistic \citep{stet96} to find sources that vary coherently in time.  We then remove 
long period variables (LPVs) and run a period-search algorithm to find periodic variables 
based on the $\sigma_{AoV}$ statistic.  We run a periodogram filter to remove 
spurious variables due to aliasing, resulting in a final set of variable star 
candidates.  Finally, we visually inspect the remaining lightcurves to remove false positives.

\subsection{Stetson $J$ Statistic} \label{sec:j}

For each of the 66,638 stars, we compute the Stetson $J$ statistic, using the implementation 
from \citet{kal98}.  This statistic identifies coherent variable stars by selecting for 
photometric variations that are correlated in time.  After visually examining a 
number of lightcurves, we define a cutoff of $J=0.7$ to select variable candidates, 
deliberately choosing a liberal cut on $J$ since we have several more tests to 
filter out non-variables.  We eliminate those stars 
with $R_K<9$, since the lightcurves of the brightest stars in our data set are 
dominated by systematics due to saturation effects.  We also eliminate any star that is less 
than 10 pixels away from stars with $R_K<9$, since extended wings and bleed trails from the bright
stars create false variability.  We finally remove any candidate 
that is less than 13 pixels away from a variable candidate with a higher RMS in 
flux.  This eliminates false positives due to constant stars close to true 
variables.  After these cuts we are left with 3430 candidate variable stars.

\subsection{Variable Clustering} \label{sec:cluster}

Among the 3430 variable candidates that pass the cuts 
described above, most of the stars are clustered around a few of the subsection boundaries, 
mostly between subsections along the field edges where the fitting parameters vary 
most, and adjacent subsections closer to the field interior.  We assume that 
all spatial clustering of variables is due to artifacts in the reduction process, and 
we suspect that the reason for the clustering has to do with how large changes in PSF 
size and shape across the field affect ISIS.
 
As described in \S \ref{sec:obs}, there was an intranight drift in the telescope 
pointing.  One effect of this drift is that stars undergo slight changes in the PSF shape 
and size during each night.  When ISIS convolves the reference image, it uses different 
parameters for each subsection (see \S \ref{sec:imsub}).  Subsections at the edges of 
the field experience the strongest optical distortions due to the wide field, and ISIS 
has the most difficulty fitting the convolution parameters in those areas.  At 
the edges of those subsections the assumptions used to compute the fitting 
parameters break down.  For stars in those areas, the intranight drift means that 
any consistently inadequate convolution will show up as photometric variability on 
timescales comparable to the drift rate.

To eliminate false variability due to this effect, we want to remove any 
candidates that appear in areas with high spatial clustering.  However, these 
areas are not precisely defined -- they result from the 
combination of the ISIS subsection grid, the direction and speed of the drift, and 
the nature of the optical distortions.  We therefore devised an algorithm to remove 
from our list any stars that are in clustered areas.  We divide the entire 
field into boxes 100 pixels on a side, and count the number of variable candidates 
in each box.  We perform this process four times, with each grid offset from the 
previous one by 20 pixels in both the X and Y directions.  We thus have four 
staggered grids with which to measure the clustering of the variable candidates.  We 
classify all candidates that appear in a box with more than two other candidates 
in any of the grids as spurious and eliminate them from our sample, leaving 1101 
variable star candidates.

\subsection{Identification of Long-Period Variables} \label{sec:lpvs}

There are some stars that pass the cut on $J$ that do not exhibit periodic 
variability.  Many are long-period variables (LPVs) that show monotonically 
increasing or decreasing brightness during our campaign, some of which may vary 
periodically but on time scales longer than our campaign.  We identify such objects and 
remove them from our later analysis, which focuses on identifying periodic variables.

\begin{figure}[b]
\epsscale{1.0}
\plotone{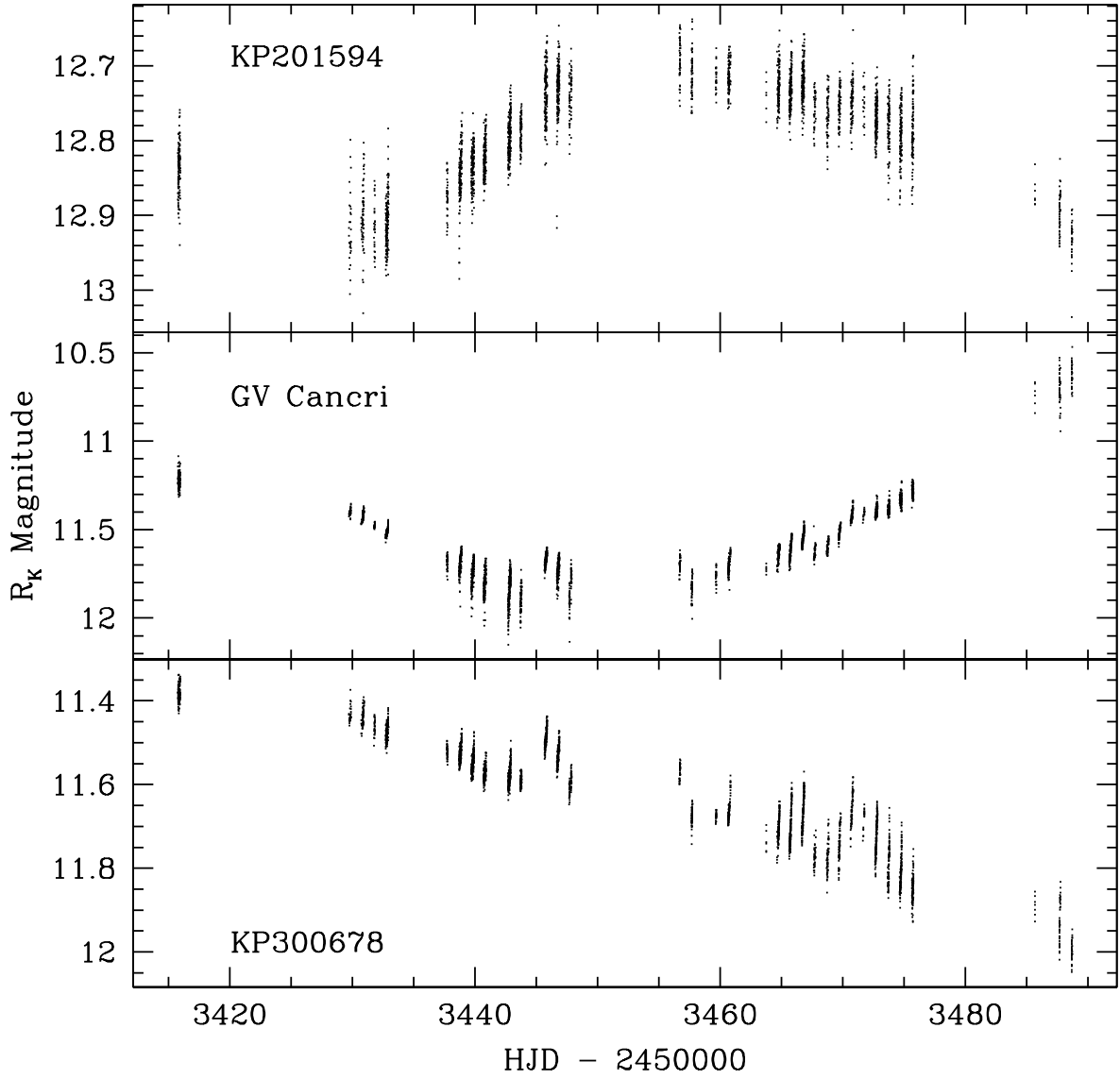}
\caption{Lightcurves of three LPVs.  We identify KP300603 as semiregular variable star GV Cnc.}
\label{fig:lpv}
\end{figure}

To identify LPVs, we use 
the method described in section 4.3 of \citet{hart04}, in which a star is defined 
as an LPV when a parabola fits the lightcurve much better than a horizontal line.
We fit a parabola to the 1,101 remaining variable candidates and calculate 
the $\chi^2/N_{\rm dof}$ for the fit, which we call $\chi^2_{N-3}$, along with 
the $\chi^2/N_{\rm dof}$ for the fit to the mean, $\chi^2_{N-1}$.  There are 52
stars for which $\chi^2_{N-3}/\chi^2_{N-1} > 0.8$, and therefore identified as 
candidate LPVs, which we eliminate from our remaining variable candidate list.  We 
show the lightcurves of three of our LPVs in Figure \ref{fig:lpv}.

A much smaller fraction of our stars are LPVs (52 out of 66,638 stars, or 0.078\%), 
than the 1,535 LPVs out of 98,000 stars, or 1.6\%, found by \citet{hart04}.  We attribute 
this difference partly to our greater observational time 
baseline (74 nights vs. 30 nights), since we end up classifying stars with 
periodic behavior in that range as regular periodic variables rather than as 
LPVs.  Also, the Praesepe field is located well out of the Galactic Plane, and we therefore 
expect many fewer background giants than in the field that was observed 
by \citet{hart04}.  Since background giants are one of the main types of LPVs (e.g. Mira 
variables), we would therefore expect fewer LPVs when observing at higher galactic latitudes.

\subsection{Periodicity Search} \label{sec:pervar}

We use the period search algorithm of \citet{sc96} to select periodic variables, adopting 
the implementation by J. Devor.  The Schwarzenberg-Czerny algorithm reports a 
periodicity likelihood statistic $AoV$, and the Devor method analyzes that information 
to estimate $\sigma_{AoV}$, a measure of the confidence of the lightcurve's periodicity.  We apply 
the algorithm to the 1,049 stars remaining in our catalog after removal of the LPVs.  

\begin{figure}[b]
\epsscale{1.0}
\plotone{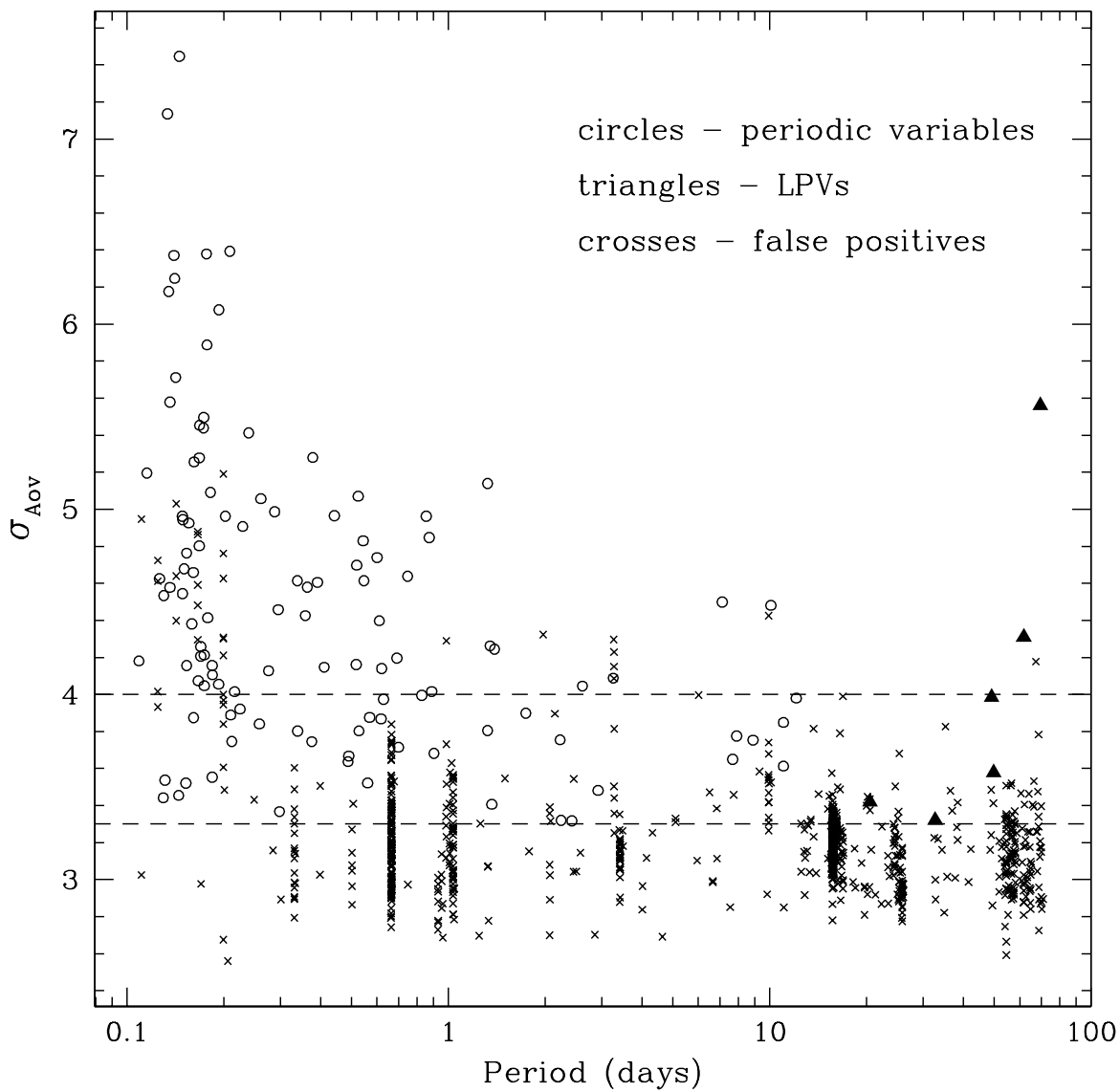}
\caption{Plot of period vs. $\sigma_{AoV}$ for the 1,049 stars left after initial removal of 
LPVs.  The dashed lines show the two cuts on $\sigma_{AoV}$.  Stars classified as periodic 
variables are marked by circles and false positives are marked by crosses.  As small number 
of stars that were not removed by the automatic filter for LPVs are determined upon visual 
inspection to be LPVs, and are marked by triangles.}
\label{fig:aov_per}
\end{figure}

The plot of the best-fit period $P$ vs. $\sigma_{AoV}$ (Figure \ref{fig:aov_per}) 
shows significant aliasing effects at periods that are integer multiples or fractions 
of 1 day.  We thus need to make further cuts to select the true variables from among our 
list of candidates.  First, we construct three histograms in log($P$).  Each histogram is 
shifted in log($P$) by $1/3$ of the width of a bin.  All objects that appear in bins in 
any of the three histograms that have more than seven objects in them are rejected.  Any 
remaining objects with $\sigma_{AoV} > 3.3$ are retained.  To ensure that true variables 
were not accidentally eliminated by the binning procedure, any stars that have 
$\sigma_{AoV} > 4.0$ are automatically retained, even if they are caught by the aliasing 
identification algorithm.  

After these cuts, we are left with 182 variable star candidates.  We then examine the lightcurves of the 
remaining candidates by hand, and find that 70 of these are false variables, all of which 
slightly missed the clustering or alias filters.  We also find that six of the candidates, all 
of which have estimated periods longer than 20 days, are in fact LPVs.

\section{Transit Search} \label{sec:trans}

The primary goal of the main KELT survey is to identify possible planetary transits.  We have 
used the commissioning data set to build our software pipeline and test our data reduction and 
analysis procedures, without expecting to identify transit candidates in this data set.  However, it 
is still useful to search for transits, since there is still a chance of discovery, and the 
transit search could also yield interesting variable stars not found through the earlier variable 
selection method.  Furthermore, we would expect to find false positives in the transit 
search -- astrophysical events that look like transits, such as low-mass stars transiting a 
solar-type star, which would have lightcurves similar to planetary transits.  Finding these events would 
demonstrate our ability to detect actual transits in the data.

In order to search the data for transits, we take two main steps.  The first is to 
apply a detrending algorithm to reduce or remove a variety of systematics.  The second is to 
run a transit search algorithm over the detrended data and identify transit candidates.

\subsection{Detrending}

Wide-field transit surveys are susceptible to a number of systematic errors.  Observing 
objects for long stretches of the night requires the telescope to cycle through significant 
changes in airmass.  A wide field of view allows differential cloud patterns to 
complicate the process of obtaining accurate relative photometry.  Temperature changes 
can affect the optics or detector performance.  With so many sources of systematic error, it 
can be prohibitive to attempt to identify, measure, and compensate for all of these 
effects.  Instead, several methods have been developed to identify all generic systematic 
effects in data sets of this type.  The method we choose to implement is the SYSREM 
algorithm \citep{tmz05,mazeh07}.  In brief, this algorithm identifies and subtracts out linear 
trends that appear in a large portion of lightcurves in a given data set.

\begin{figure}[b]
\epsscale{1.0}
\plotone{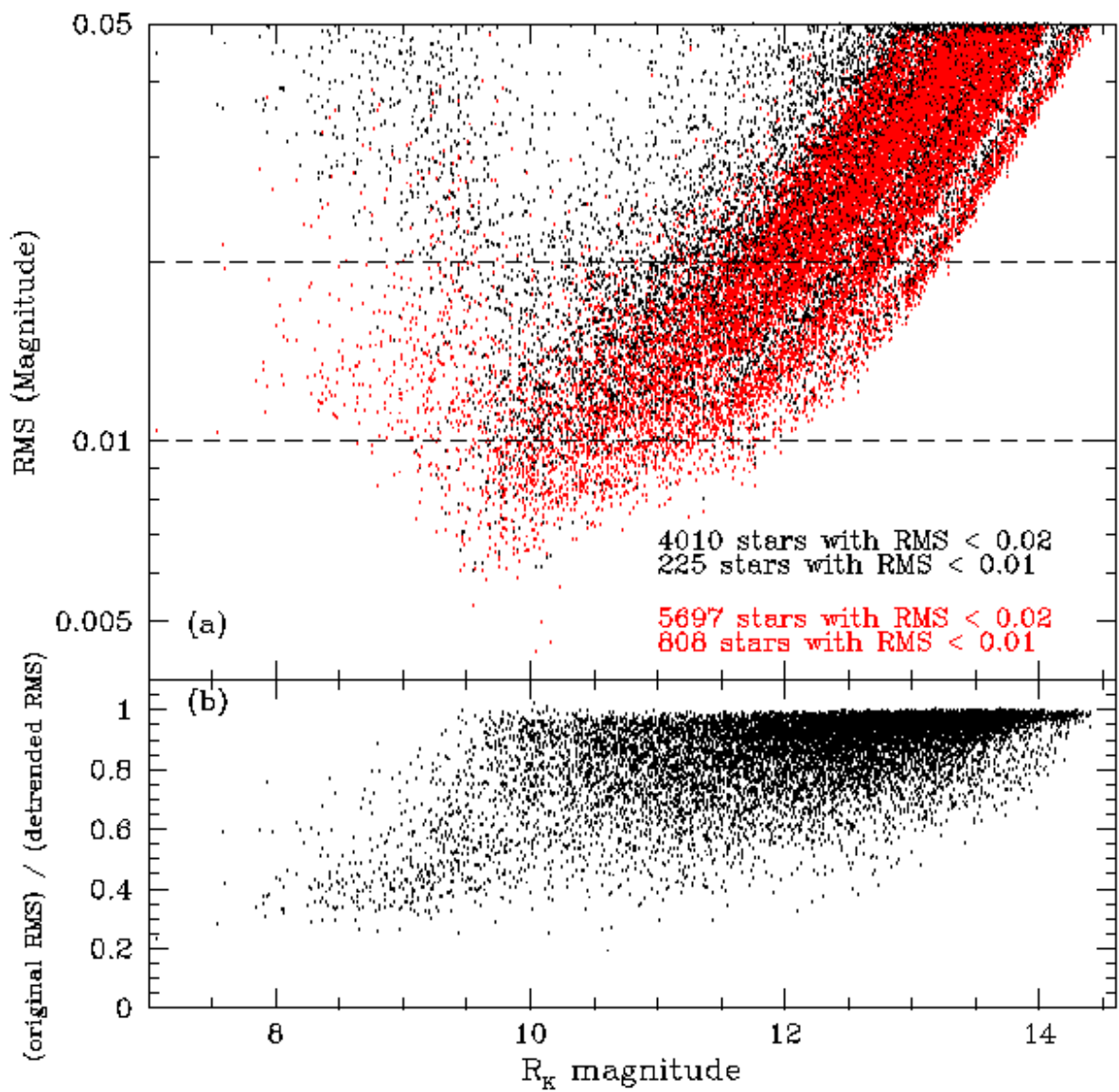}
\caption{Panel (a) shows the RMS plot for 15,012 stars before (black) and after (red) application of 
SYSREM detrending.  Structure in the locations of the fainter stars results from the 
separate reductions of the four regions to compensate for the changes in the FWHM pattern. Panel (b) 
shows the change in RMS by magnitude.}
\label{fig:sys}
\end{figure}

We apply SYSREM to the KELT lightcurves that have been reduced using the procedures 
described in \S \ref{sec:datar}.  Since lightcurves with very large variations can create 
problems for detrending programs, and are not useful for the identification of low 
level systematics, we only apply SYSREM to the 15,012 stars from our data set 
with RMS $< 5\%$.  We recalculate the RMS for the detrended lightcurves and display 
the results in Figure \ref{fig:sys}.  For most of the lightcurves, detrending improves 
the RMS by about $\sim10\%$, although for a few stars, especially towards the 
bright end, the RMS improves by a factor of 3 to 4.

\begin{figure}[b]
\epsscale{1.0}
\plotone{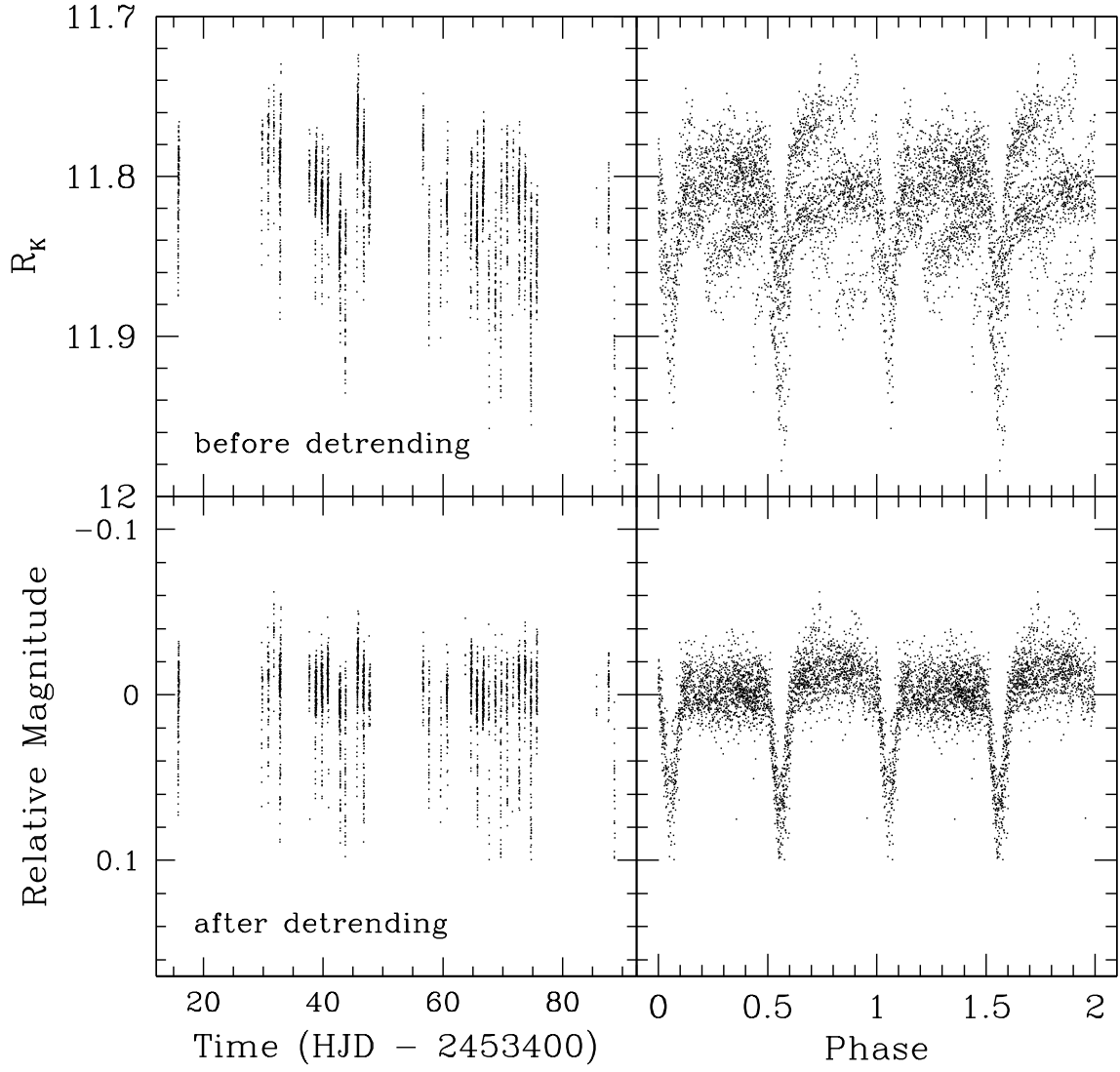}
\caption{Lightcurves of the detached eclipsing binary KP102511 (Period=0.5584 days) before and after detrending 
with SYSREM.  The value of $\sigma_{AoV}$ increased from 3.1 to 4.8.}
\label{fig:sysex}
\end{figure}

We suspect that the reason that detrending does not improve the RMS to a 
greater degree is related the the FWHM problems described in \S \ref{sec:focus}.  Algorithms 
like SYSREM are designed to remove linear systematics, but it is probable that 
the systematic errors caused by the FWHM changes are higher order, and so are not 
rectifiable by linear detrending.  Even so, detrending manages to improve the RMS by a small 
amount, and in some cases works very well, as shown in Figure \ref{fig:sysex}.  For the periodic 
variables we identify with non-detrended RMS below 5\%, detrending improves $\sigma_{AoV}$ on average by 10\%.

\subsection{Transit Selection} \label{sec:transel}

To detect transits, we apply a version of the Box-Fitting Least-squares (BLS) 
transit-search algorithm \citep{kovacs02} implemented \citet{burke06}.  This algorithm 
cycles through a range of periods and phases searching for a transit-like event 
and selects the one with the highest significance.  For each lightcurve, we calculate 
the $\Delta \chi^2$ between a constant flux and the best-fit transit model, 
the $\Delta \chi^2_-$ for the best-fit antitransit (a brightening rather than 
dimming), the fraction of $\Delta \chi^2$ that results from a single night $f$, and 
the transit period $P_t$.  We then examine five parameters: $\Delta \chi^2$, 
$\Delta \chi^2 / \Delta \chi^2_-$, $f$, $P_t$, and the transit depth.  We also 
examine Digital Sky Survey (DSS) images of the stars which have much higher resolution 
than KELT, to check whether the single KELT sources consist of blends of more than one star.

Because of the unique characteristics of this data set, we have not elected to 
construct a rigorous transit selection algorithm using cuts.  Instead we sort the 
lightcurves based on each of the five parameters described above, and examine by 
eye the 100 best lightcurves identified by each criterion.  We classify the 
lightcurves as either possible transits, variables, or neither.  We find four stars 
with transit-like behavior that are unblended and that are identified with known 
2MASS stars.  We also find 38 additional variable stars that were not found with 
the steps described in \S \ref{sec:vars}.  Of those objects, 31 had $J<0.7$, which 
is not surprising since the transit-like signals that BLS searches for would not 
necessarily show the coherent variations characteristic of a variable star with 
a large $J$ value.  Of the seven other stars, one was eliminated by the 
cluster-rejection routine, and the remaining six had $\sigma_{AoV}<3.3$.  

\section{Variable Stars} \label{sec:vcat}

After these procedures we are left with 208 variables, which include 
the 52 LPVs identified in \S \ref{sec:lpvs}, the 6 LPVs found 
in \S \ref{sec:pervar}, the 112 periodic variables found in \S \ref{sec:pervar}, and 
the 38 periodic variables found in \S \ref{sec:transel}.

\subsection{Matching to Known Variables}

There are 168 known variables within our field of view combining
The General Catalog of Variable Stars \citep[GCVS4.2;][]{sd04} and the 
New Catalog of Suspected Variable Stars \citep[NSV;][]{ku82}, 

The magnitudes of the variables are reported in 
either Johnson $V$ or $U$, or photographic magnitude $p$.  The range of magnitudes for 
which we found variables in the KELT data using all the methods described above 
are $9 < R_K < 16$.  Out of the 168 variables, 72 have reported magnitudes outside that 
range, and an additional 17 are within $15 < V < 16$.  We only try to 
match known variables with reported magnitudes between 9 and 15. 

We check whether we detect the remaining 79 variables in our data.  We 
search for KELT counterparts using a matching radius of two 
pixels (19\arcsec) and identify 63 matches.  Of the 16 known variables we 
do not detect, three appear either saturated or blended with saturated stars in 
our data, and 13 do not appear to have counterparts 
within two pixels of the reported positions, suggesting that the reported 
positions, proper motions, or magnitudes may be incorrect.

We then compare our list of 208 variables with the 63 detected known variables and 
find 14 matches.  Of the 49 known variables we do not classify as variable through 
our tests, five have counterparts that are either brighter than $R_K=9$ or 
blended with a bright star.  Another 38 are irregular, eruptive, or unspecified 
variables, which we would not expect to detect with our periodicity-based 
search algorithms.  Three are eclipsing variables of unspecified type with no periods 
listed in the catalogs.  Upon investigation of their KELT counterparts, no eclipses are seen, 
indicating that either our observations missed the eclipses, or that the eclipse 
depths were too small to appear in our data.  One is listed as an RR Lyrae variable 
with no period given, for which the KELT lightcurve shows no periodic variation.  The 
remaining two are eclipsing binaries which are removed in the clustering filter, but 
can be seen in the KELT lightcurves.  We add those two stars to our list of variables, giving a total of 210.

We thus have KELT lightcurves for 16 known variables.  We find that one KELT LPV is the 
semi-regular variable GV Cancri.  The lightcurve for this object is shown in Figure \ref{fig:lpv}.  We 
find three RR Lyrae variables, CQ Cancri, AN Cancri, and EZ Cancri, the 
first two of which have periods reported in the catalogs that we confirm.  One more variable, EF Cancri, is 
listed as a W UMa contact eclipsing binary, but the lightcurve, despite the low-quality 
photometry, appears to indicate that it, too, is an RR Lyrae variable.

We detect six known eclipsing binaries: EH Cancri, GW Cancri, FF Cancri, RU Cancri, 
NSV 04207, and NSV 04158, in 
addition to the two eclipsing binaries that were caught by the clustering filter: TX Cancri and RY 
Cancri.  We confirm the reported periods of RY Cancri and TX Cancri, but we 
find that RU Cancri has a period of 10.0591 days rather than the reported period of 
10.172988 days.  The remaining variable stars do not have previously reported periods.

The star NSV 04269 is listed as a semiregular variable in the NSV catalog, but our 
lightcurve shows it to be an eclipsing binary.  The star NSV 04069 is not listed 
as any variable type; we classify it as an eclipsing binary.  Lastly, the star FR Cancri is 
listed as a BY Draconis variable, consistent with our lightcurve.

\subsection{Variable Star Catalog}

We list the properties of our variable stars in Table \ref{tab:var}.  For each star we list the KELT ID 
number, mean $R_K$ magnitude, coordinates in Right Ascension and Declination (J2000.0), $JHK$ colors from 
2MASS and 2MASS ID for those that matched to 2MASS objects, period (for non-LPVs) in days, type of variable 
based on our classification, and the GCVS or NSV ID and classification for those that we have 
identified with previously known variables.

\begin{figure}[b]
\epsscale{1.0}
\plotone{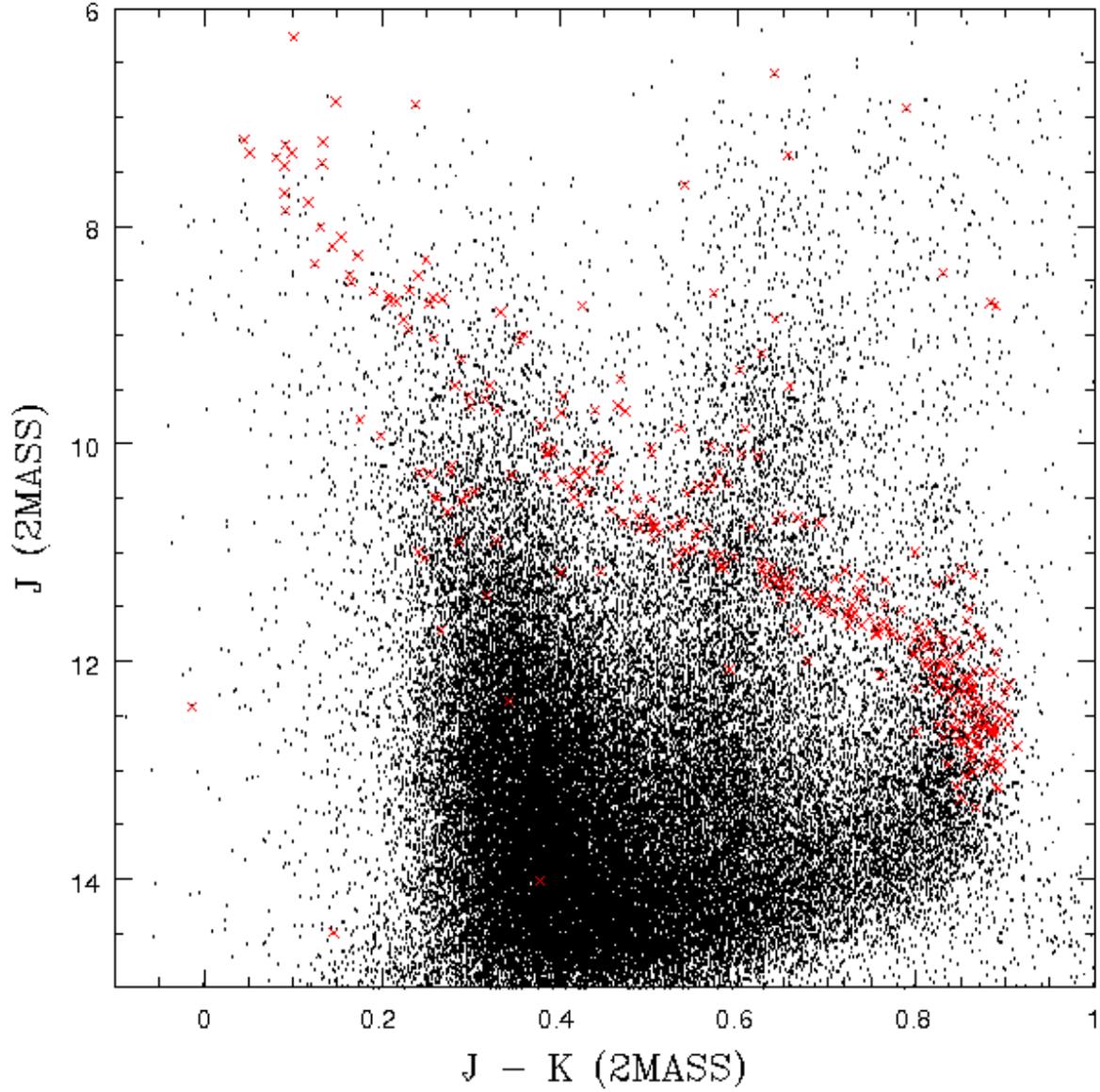}
\caption{Color-magnitude diagram for cluster stars and field stars.  Only field stars that match 
to a single 2MASS source are plotted; all cluster stars have unique 2MASS matches.  Red crosses 
mark stars listed as members of Praesepe by WebDA.}
\label{fig:cmd1}
\end{figure}

\begin{figure}[b]
\epsscale{1.0}
\plotone{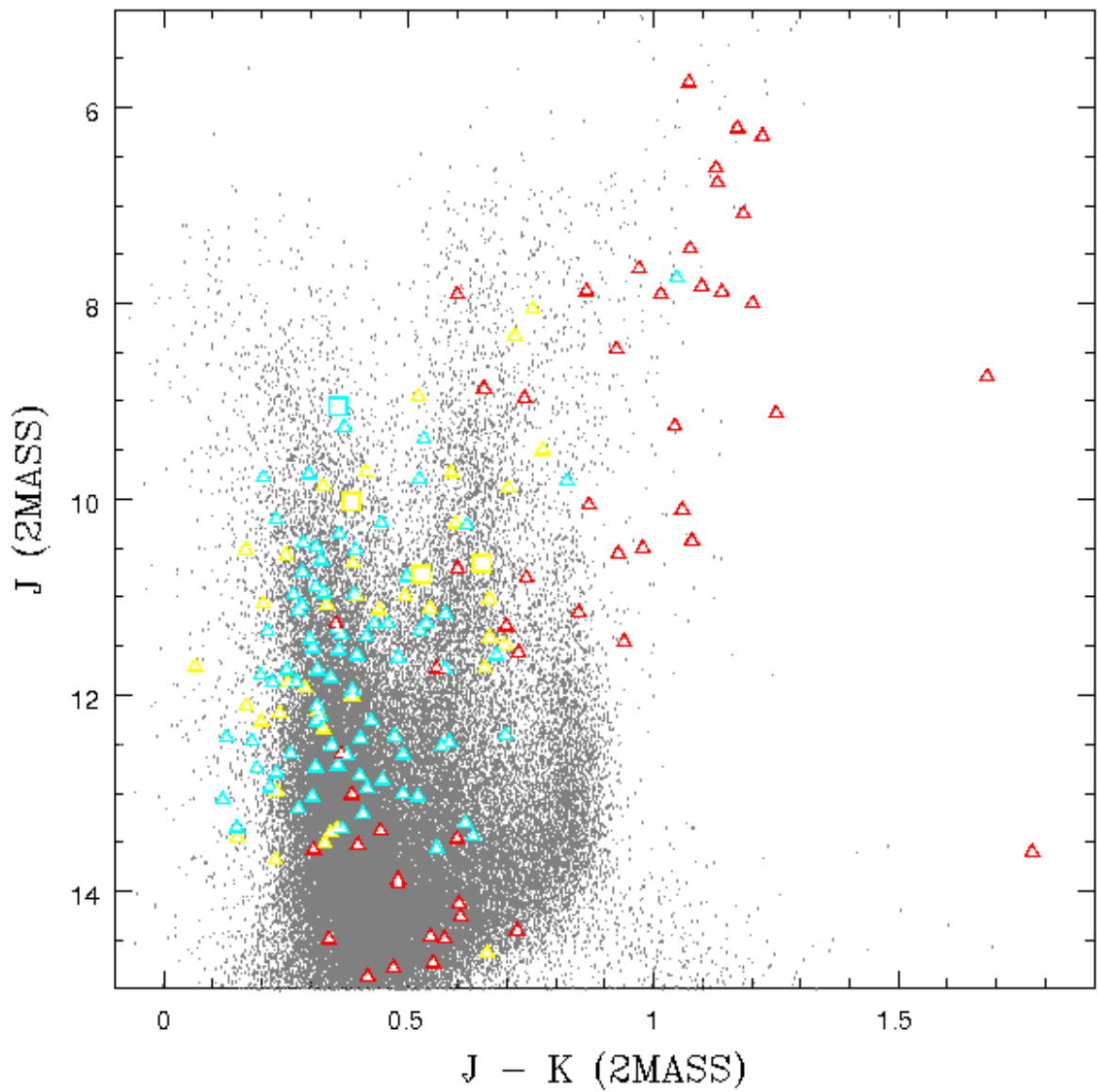}
\caption{Color-magnitude diagram for variable and non-variable stars.  Triangles plot the positions of field 
variables, while squares plot the four cluster variables.  Red symbols mark LPVs, blue symbols mark eclipsing 
binaries, and yellow symbols mark pulsating variables.  Only non-variable stars that match to a single 2MASS 
source are plotted as dots.  The eight variable stars that match to more than one 2MASS source are plotted according to the $J$ 
and $K$ colors of the nearest 2MASS source.  The 27 variables that do not match to any 2MASS source are not plotted.}
\label{fig:cmd2}
\end{figure}

We plot the 2MASS infrared color-magnitude diagram (CMD) for all our stars in Figures \ref{fig:cmd1} 
and \ref{fig:cmd2}.  In Figure \ref{fig:cmd1} we show Praesepe cluster members along with all 
background stars.  The cluster members form a coherent main sequence.  The background stars form 
three populations in $J-K$ color.  The largest group, at $J-K \sim 0.4$ consists of main 
sequence stars.  The group at $J-K \sim 0.6$ consists of red giants, and the last group, 
at $J-K \sim 0.8-0.9$, consists of nearby late-type stars, clearly overlapping the same stellar 
population at the low end of the Praesepe cluster. 

\begin{figure}[b]
\epsscale{1.0}
\plotone{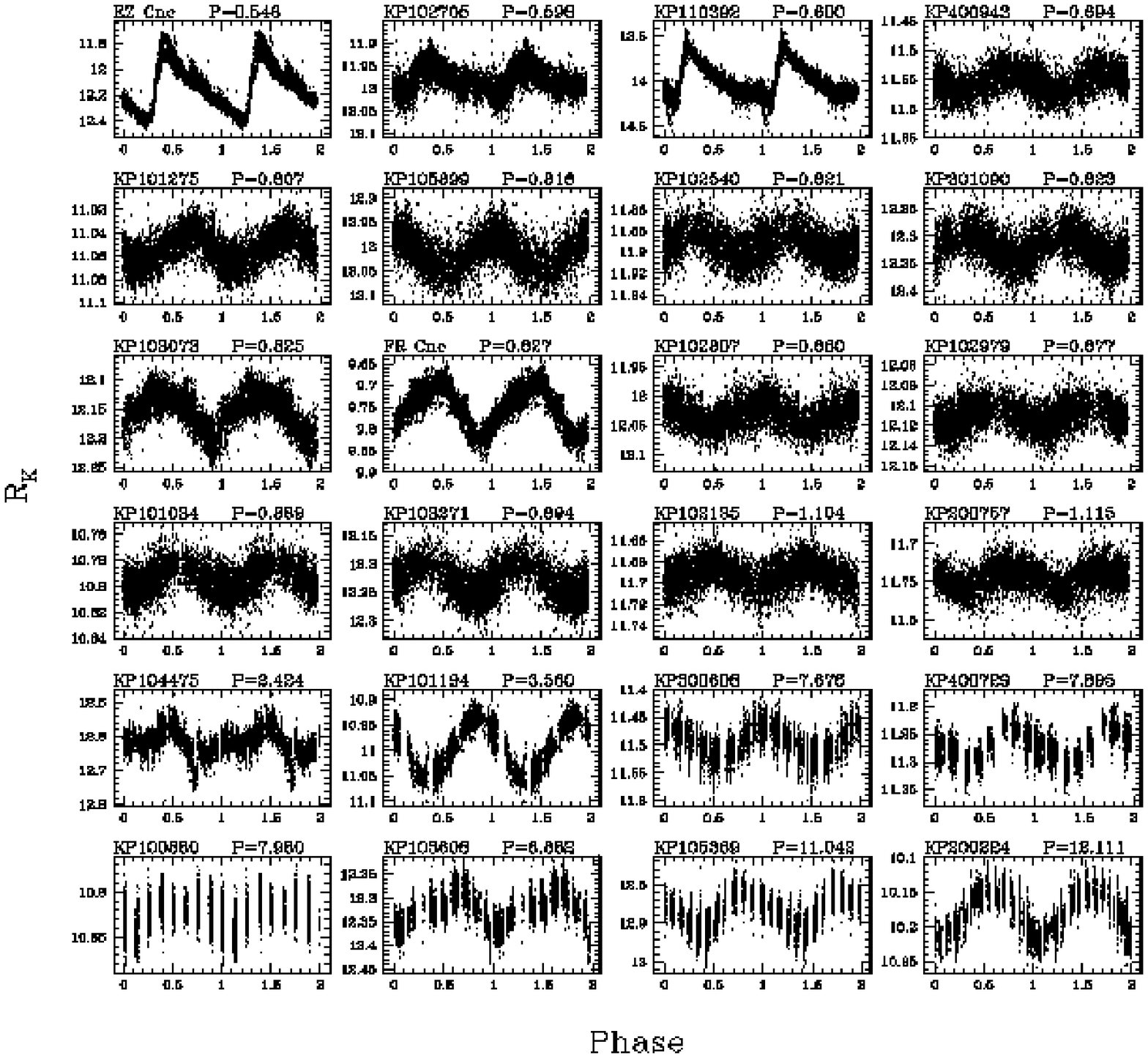}
\caption{Lightcurves of pulsating periodic variables identified by KELT, sorted by period.}
\label{fig:var.puls2}
\end{figure}

\begin{figure}[b]
\epsscale{1.0}
\plotone{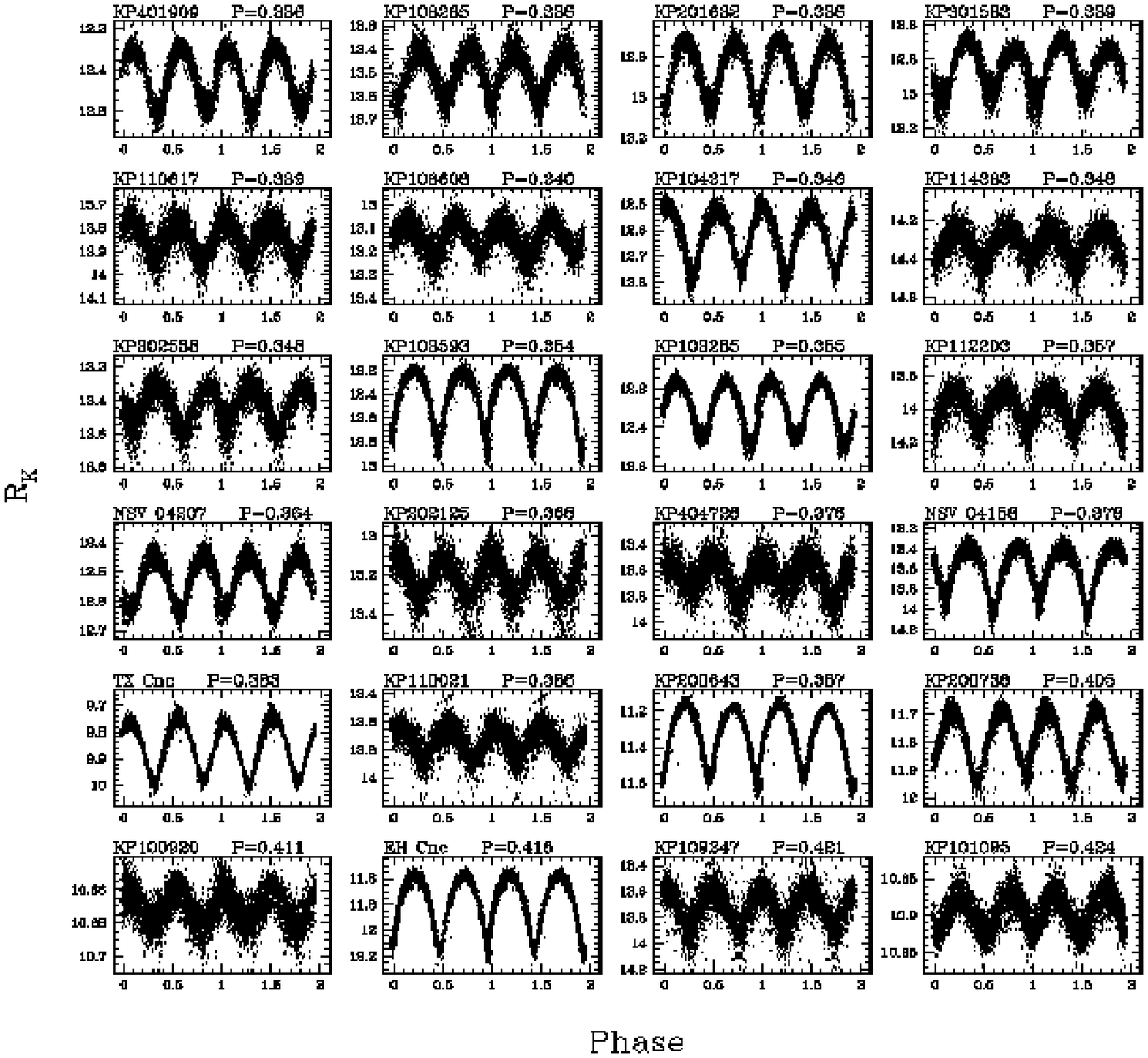}
\caption{Lightcurves of eclipsing periodic variables identified by KELT, sorted by period.}
\label{fig:var.eb2}
\end{figure}

\begin{figure}[b]
\epsscale{1.0}
\plotone{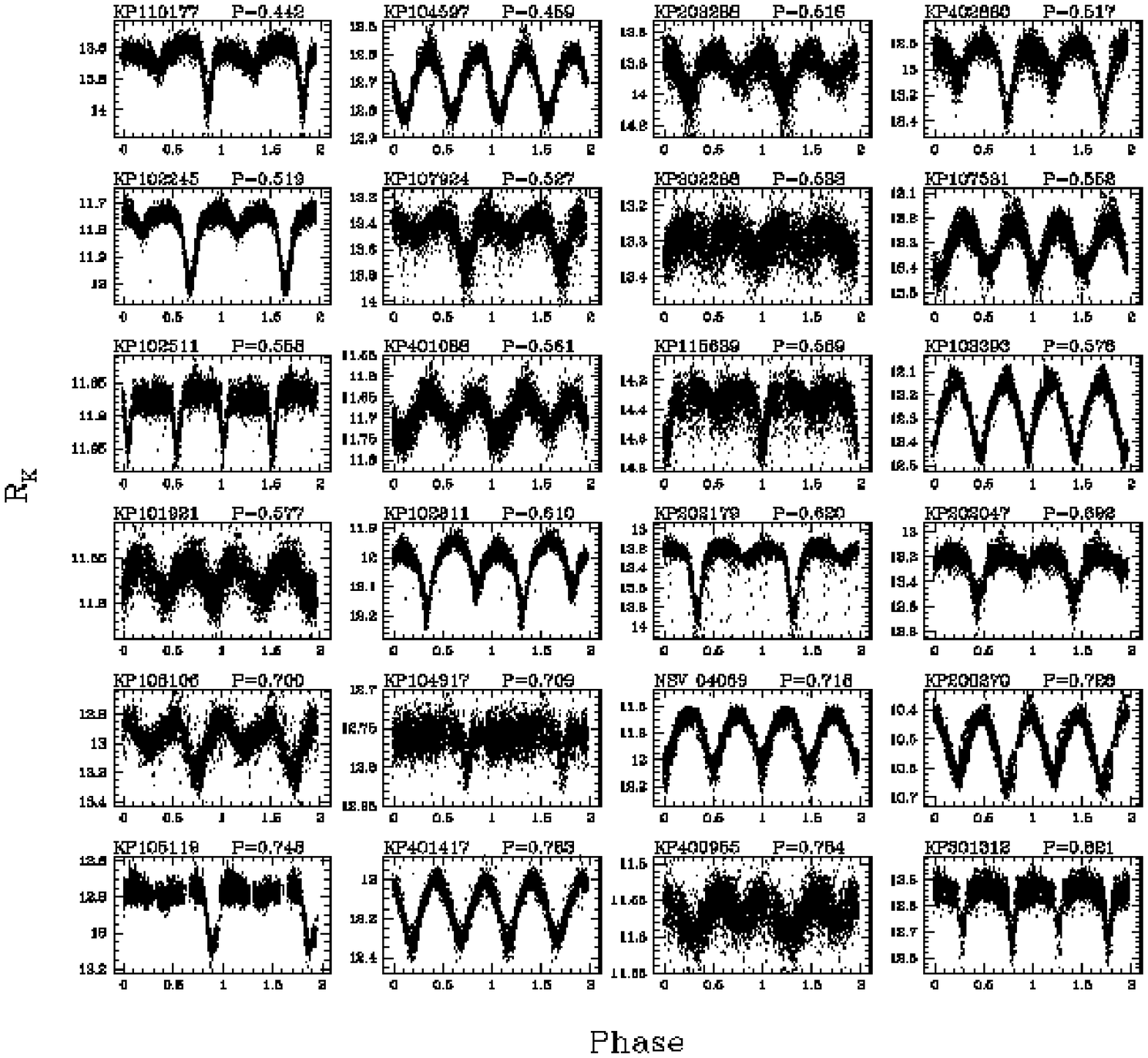}
\caption{Lightcurves of eclipsing periodic variables identified by KELT, sorted by period.}
\label{fig:var.eb3}
\end{figure}

\begin{figure}[b]
\epsscale{1.0}
\plotone{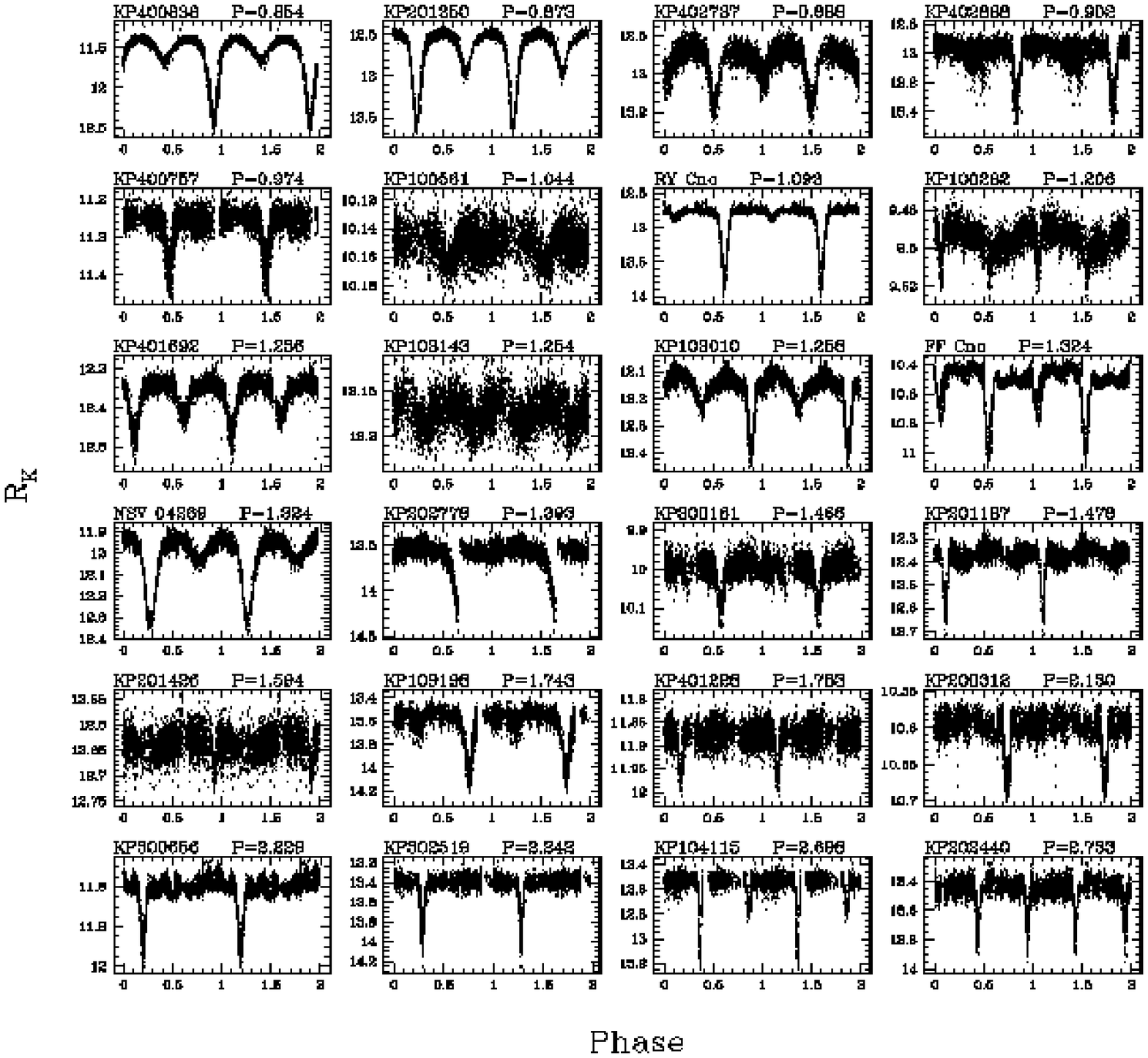}
\caption{Lightcurves of eclipsing periodic variables identified by KELT, sorted by period.}
\label{fig:var.eb4}
\end{figure}

\begin{figure}[b]
\epsscale{1.0}
\plotone{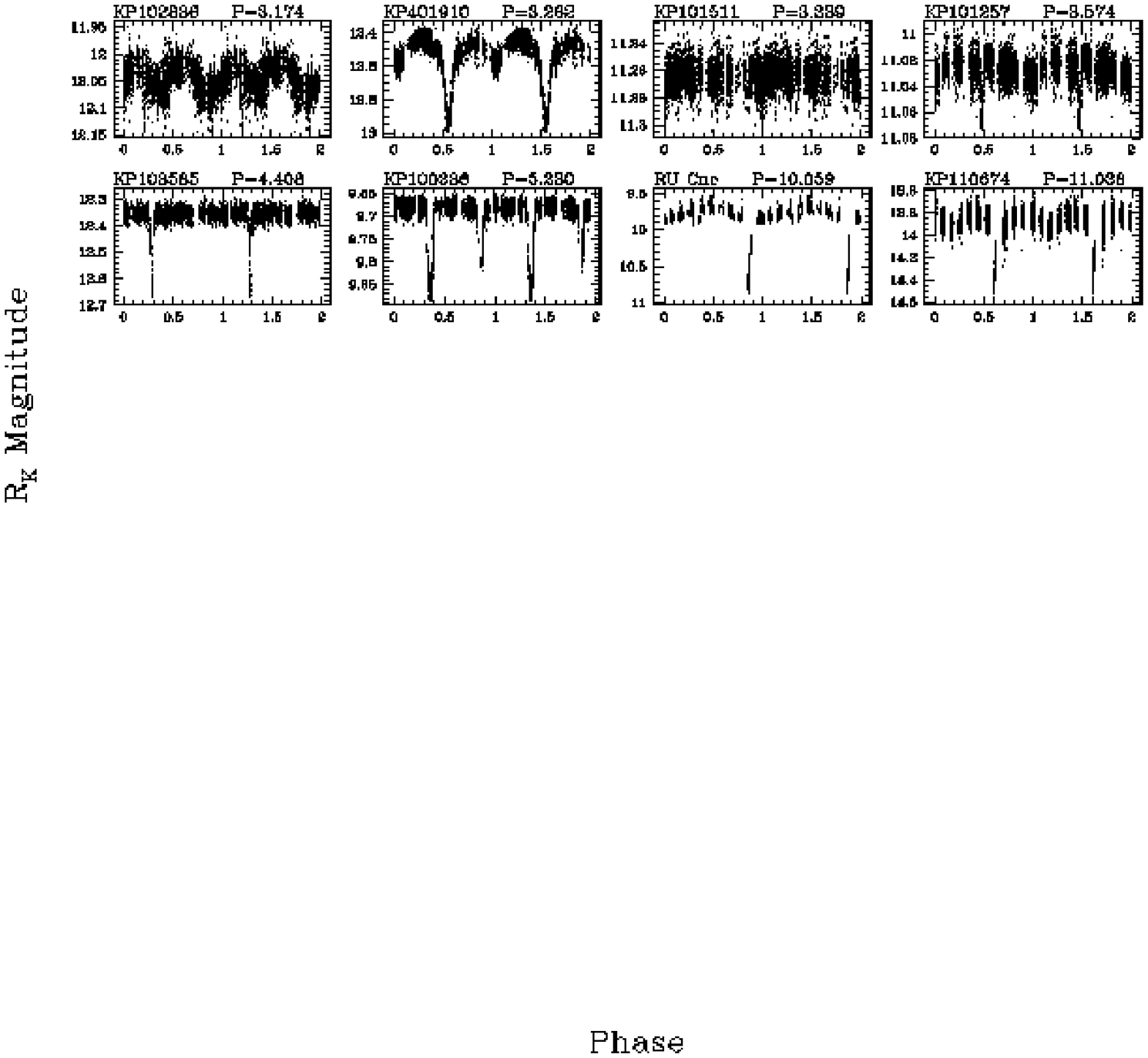}
\caption{Lightcurves of eclipsing periodic variables identified by KELT, sorted by period.}
\label{fig:var.eb5}
\end{figure}

We plot the lightcurves of the periodic variables we identify with the methods described above, 
along with the two previously known periodic variables that missed our cuts.  We classify 48 
variable as pulsators, and we plot their lightcurves in Figures \ref{fig:var.puls1} and 
\ref{fig:var.puls2}.  We classify 105 variables as eclipsing binaries, and we plot their light 
curves in Figures \ref{fig:var.eb1} through \ref{fig:var.eb5}.  

\section{Transit Candidates} \label{sec:tcan}

\begin{figure}
\epsscale{1.0}
\plotone{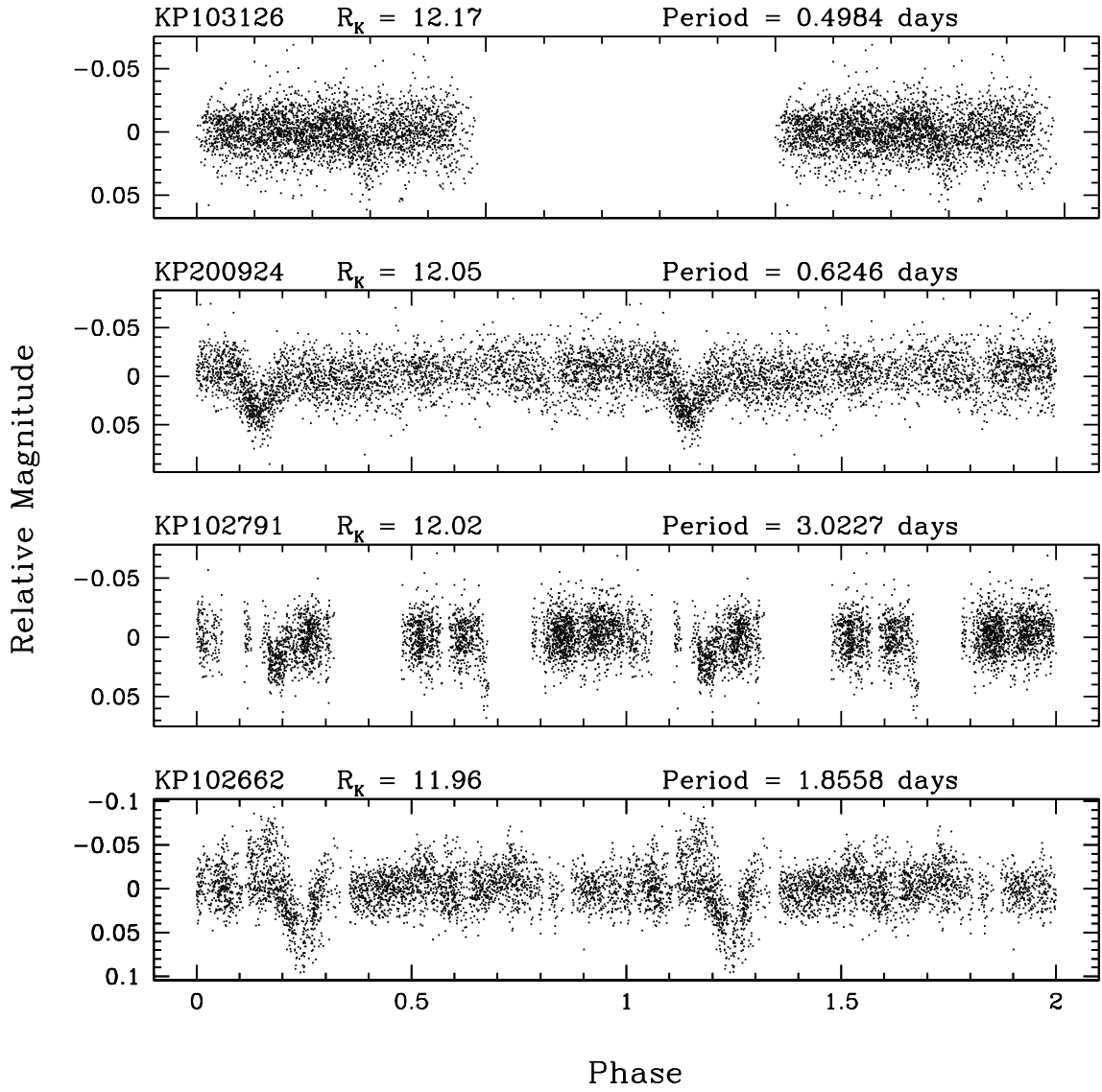}
\caption{Lightcurves of the four best transit candidates.}
\label{fig:tcands}
\end{figure}

Using the selection process from \S \ref{sec:transel}, we identify four possible transit 
candidates.  The properties  of the four candidates are listed in Table \ref{tab:tprop}, and 
their lightcurves are shown in Figure \ref{fig:tcands}.  These four lightcurves all show 
events with depths of 5\% or less and are not blended with nearby stars in DSS images.

Unfortunately, because of the time that has elapsed between the original observations and the 
final transit selection, calculations of the eclipse ephemerides for the transit candidates 
are too inaccurate for targeted photometric follow-up.  The periods we derive for the 
candidates are generally accurate to of order 20 seconds, and since the original observations 
were taken over two years before the candidates were identified, we do not have ephemerides 
accurate to within an hour.  We therefore use spectroscopic observations to rule out 
astrophysical false positives.

\subsection{Spectroscopic Follow-Up}

The four transiting-planet candidates reported in \S \ref{sec:transel}
were followed up spectroscopically using the CfA Digital Speedometer
\citep{Latham1992} on the 1.5-m Tillinghast Reflector at the F.\ L.\
Whipple Observatory atop Mount Hopkins, Arizona.  This instrument has
been used extensively for the initial spectroscopic reconnaissance of
transiting-planet candidates identified by Vulcan \citep{Latham2003}
and by TrES and HAT \citep{Latham2007}.  Single-order echelle spectra
centered at $5187$ \AA\, were recorded using an intensified
photon-counting Reticon detector with a spectral resolution of 8.5
\kms\ and typical signal-to-noise ratio of 10 to 20 per resolution
element.  After rectification to intensity versus wavelength, the
observed spectra were correlated against extensive grids of synthetic
spectra drawn from a library calculated by Jon Morse using
\citet{Kurucz1992} model atmospheres and codes.  This allowed us to
estimate the effective temperature and surface gravity of the star,
assuming solar metallicity, as well as the rotational and radial
velocities.  For the initial spectroscopic reconnaissance we normally
obtain at least two spectra, so that we can look for velocity
variations down to the level of about 1 \kms\ for slowly-rotating
solar-type stars.

In the case of KP200924 the first CfA observation revealed that this
candidate has a composite spectrum.  Plots of the one-dimensional
correlation functions clearly show two peaks corresponding to the two
stars in a double-lined spectroscopic binary, with a velocity
separation of about 200 \kms\ and rotational broadening of about 65
\kms.  This must be a grazing eclipsing binary.

\begin{figure}
\epsscale{1.0}
\plotone{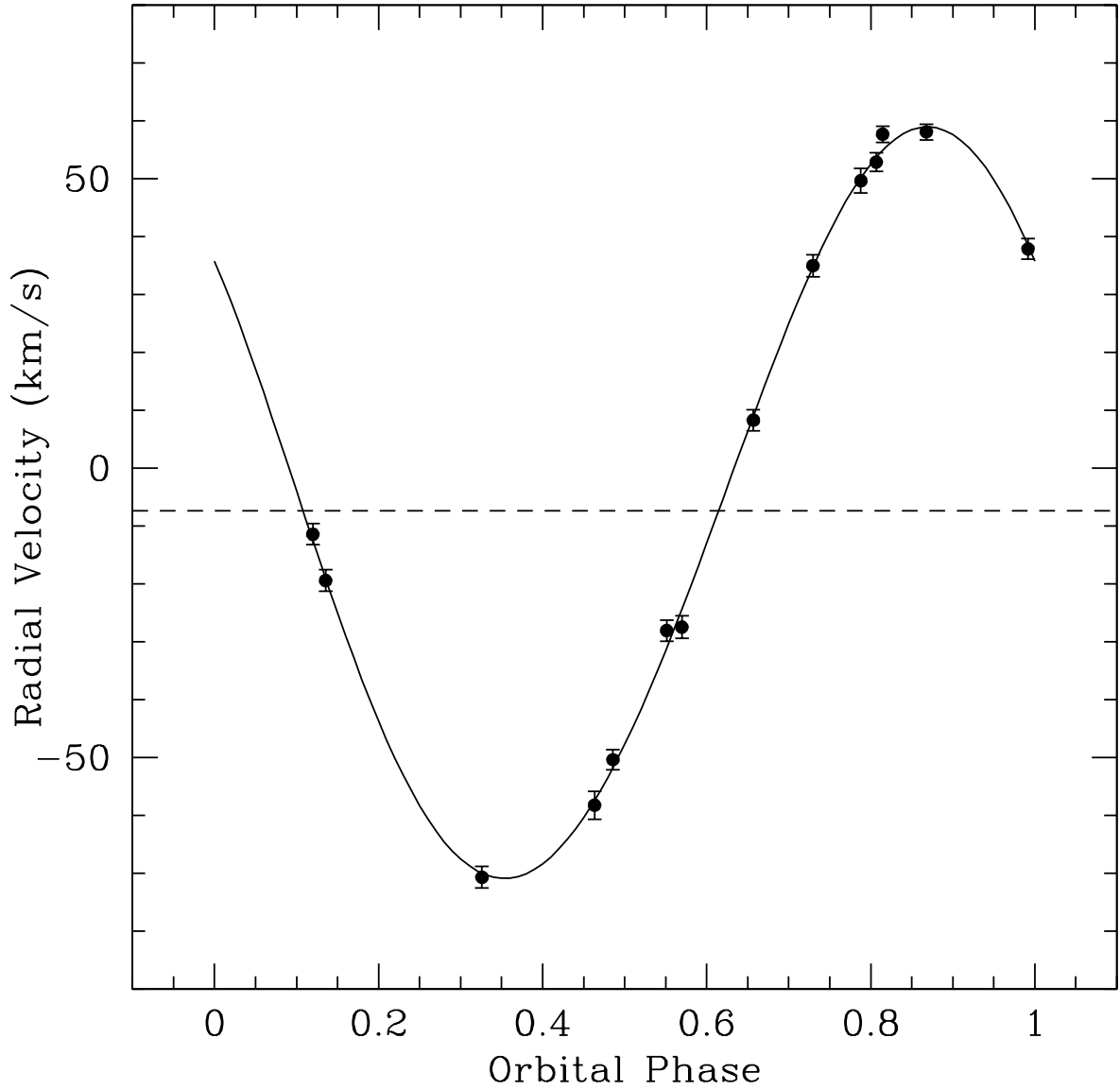}
\caption{Spectroscopic orbit of transit candidate KP102791, with a period of $3.0227 \pm 0.0011$
days.  The amplitude of the orbit clearly 
indicates that the companion is a stellar companion and not a planet.  This object turns out to 
be an eclipsing binary with a 1.5$M_\odot$ primary with a 0.75$M_\odot$ secondary.}
\label{fig:spec_curve}
\end{figure}

For KP102791 we obtained 14 CfA spectra spanning 88 days. The
spectroscopic analysis yielded an effective temperature of $T_{\rm
eff} = 7000$ K, a surface gravity of $\log (g) = 4.5$ cm sec$^{-2}$,
and rotational velocity of $V_{\rm rot} = 27.5$ \kms.  The 14
radial velocities, listed in Table \ref{tab:radvel}, allowed us to derive
a single-lined spectroscopic orbit with period $P = 3.0227 \pm 0.0011$
days, eccentricity $e = 0.033 \pm 0.026$, orbital semi-amplitude $K =
64.9 \pm 1.0$ \kms, and mass function $f(m) = 0.00856 \pm 0.00039$
solar masses, see Figure \ref{fig:spec_curve}.  Notice that the spectroscopic period is twice the
photometric period, which often happens when the secondary eclipses
look similar to the primary eclipses.  The eccentricity is
indistinguishable from circular, suggesting that the orbit has been
circularized by tidal forces.  Thus, it is not unreasonable to assume
that the rotation of the two stars has been synchronized and aligned
with the orbital motion.  In this case the observed spectroscopic line
broadening can be used to estimate the radius of the primary star,
which comes out to about 1.6 solar radii.  This in turn implies that
the primary has not evolved very much, which is consistent with the
surface gravity derived from the spectra. Adopting a mass of 1.5 solar
masses for the primary, the mass of the unseen secondary implied by the
mass function is about 0.75 solar masses.  The ephemeris for future
primary eclipses based on just the radial velocity data is
$2454170.445 \pm 0.069 + (3.0227 \pm 0.0011) \times E$.  Observations with
KeplerCam on the 1.2-m reflector at the Whipple Observatory revealed
an ingress starting at heliocentric Julian date 2454165.84.
Unfortunately the event started six hours later than predicted by
the photometric ephemeris available at that time from data that were
already two years old, and the ingress was still in progress and
already 0.025 magnitudes deep when the telescope reached its pointing
limits.  It turns out that this particular transit event must have been
a secondary eclipse, with its center almost exactly 1.50 cycles
after the spectroscopic epoch for primary eclipses quoted above.

For KP102662 we obtained five CfA spectra spanning 33 days.  The
spectroscopic analysis yielded $T_{\rm eff} = 6500$ K, $\log (g) = 4.5$
$V_{\rm rot} = 3.5$ \kms, and mean radial velocity $<V_{\rm rad}> = -20.83
\pm 0.37$  \kms RMS.  The $\chi^2$ probability that the observed velocity
residuals are consistent with Gaussian errors and constant velocity is
$P(\chi^2) = 0.82$, so if the transit-like lightcurve observed for
the visible star in this system is caused by an orbiting companion,
the companion mass must be less than just a few Jupiter masses.  On
this basis KP102662 survives as a viable transiting planet candidate.
However, the transit lightcurve looks V-shaped and rather too deep to
allow a Jupiter-sized planet.  Nevertheless, this candidate deserves
further follow-up observations.  Obtaining a high-quality lightcurve
would probably prove to be time consuming, because the ephemeris is
based on data that are already two years old, so highly precise radial
velocities may be the best way to proceed.

For KP103126 we obtained six CfA spectra spanning 58 days. The
spectroscopic analysis yielded $T_{\rm eff} = 6250$ K, $\log (g) = 4.5$
cm sec$^{-2}$, $V_{\rm rot} = 0.5$ \kms, $<V_{\rm rad}> = -11.49 \pm 0.79$ \kms
RMS, and $P(\chi^2) = 0.035$.  The velocity residuals are a bit larger
than expected, but still consistent with an orbiting companion that is
no more than several Jupiter masses.  Thus this candidate also
survives as a viable transiting-planet candidate, one that might
reward highly precise radial-velocity observations.

\section{Conclusions} \label{sec:disc}

This paper has presented an analysis of the KELT commissioning data, 
consisting of a 74-day campaign towards the Praesepe open cluster.  We 
obtained lightcurves for over 66,000 stars, and identified 210 variable 
stars, of which 194 were not previously known as variable.

We have also searched for planetary transits, finding four transit 
candidates.  Follow-up observations have ruled out two of the candidates 
as being non-planetary in origin, while two remain as possible planetary 
systems.  This data set has served as the testbed for developing the variable 
and transit search algorithms that will be used to analyze data from the 
main KELT survey, and has demonstrated the ability of KELT to detect signals 
at the level of precision of transiting planets.

\acknowledgments 

We would like to thank the many people who have helped with this research, including Scott Gaudi, Mark 
Trueblood, and Pat Trueblood.  We would also like to thank Marc Pinsonneault and Deokkeun An for discussions 
about cluster and stellar properties.  We would like to thank the authors of several programs used for 
this research, including W. Pych for the program to determine FWHM, Christopher Burke for the 
implementation of the BLS algorithm, and G. Pojmanski for the {\tt lc} program for analyzing 
lightcurves.  This work was supported by the National Aeronautics and Space Administration 
under Grant No. NNG04GO70G issued through the Origins of Solar Systems program, and from 
the Kepler Mission under NASA Cooperative Agreement NCC-1330 with the Smithsonian Astrophysical
Observatory.
This publication makes use of data products from the Two Micron All Sky Survey, which is a joint 
project of the University of Massachusetts and the Infrared Processing and Analysis Center/California 
Institute of Technology, funded by the National Aeronautics and Space Administration and the National 
Science Foundation.

\clearpage

\begin{landscape}
\begin{deluxetable}{lrrrrrrlrrrr}
\tabletypesize{\footnotesize}
\tablecaption{Variable Stars Identified in KELT Observations of Praesepe\label{tab:var}}
\tablewidth{0pt}
\tablehead{  \colhead{KELT} & \colhead{$R_K$} & \colhead{RA} & \colhead{Dec} & \colhead{$J$} & \colhead{$H$} & \colhead{$K$} & \colhead{2MASS} & \colhead{Period} & \colhead{KELT} & \colhead{GCVS/NSV} & \colhead{GCVS/NSV} \\ \colhead{ID \#} & \colhead{} & \colhead{(J2000.0)} & \colhead{(J2000.0)} & \colhead{} & \colhead{} & \colhead{} & \colhead{ID \#} & \colhead{(days)} & \colhead{Class.} & \colhead{ID} & \colhead{Class.} }
\startdata
\hline
        KP100169                  &  9.136 & 133.01755 & 14.58464 &  6.212  &  5.349  &  5.041  & J08520421+1435047         	     &  \nodata &  LPV &    \nodata &    \nodata \\
        KP100282                  &  9.495 & 131.11816 & 20.20117 &  9.291  &  8.881  &  8.832  & J08442835+2012042\tablenotemark{b} &  1.2065  &   EB &    \nodata &    \nodata \\
        KP100305                  &  9.605 & 133.03393 & 21.84911 &  7.440  &  6.605  &  6.366  & J08520814+2150567         	     &  \nodata &  LPV &    \nodata &    \nodata \\
        KP100306                  &  9.606 & 131.14550 & 16.28497 &  7.646  &  6.826  &  6.675  & J08443491+1617058         	     &  \nodata &  LPV &    \nodata &    \nodata \\
        KP100336                  &  9.686 & 132.50840 & 17.87422 &  9.368  &  8.983  &  8.838  & J08500201+1752271\tablenotemark{b} &  5.2301  &   EB &    \nodata &    \nodata \\
        KP100445\tablenotemark{a} &  9.930 & 130.00713 & 18.99986 &  9.053  &  8.767  &  8.698  & J08400171+1859594         	     &  0.3828  &   EB &     TX Cnc &      W UMa \\
        KP100561                  & 10.149 & 130.99077 & 16.71554 &  7.741  &  6.932  &  6.694  & J08435778+1642559         	     &  1.0438  &   EB &    \nodata &    \nodata \\
        KP100626                  & 10.274 & 135.20934 & 18.27620 &  6.291  &  5.470  &  5.069  & J09005024+1816343         	     &  \nodata &  LPV &    \nodata &    \nodata \\
        KP100722                  & 10.408 & 131.77529 & 21.03571 &  8.869  &  8.321  &  8.215  & J08470606+2102085         	     &  \nodata &  LPV &    \nodata &    \nodata \\
        KP100880                  & 10.625 & 135.54936 & 16.38090 &  9.720  &  9.387  &  9.308  & J09021184+1622512         	     &  7.9795  & Puls &    \nodata &    \nodata \\
        KP100886                  & 10.630 & 131.78774 & 23.84970 & \nodata & \nodata & \nodata &           \nodata          	     &  \nodata &  LPV &    \nodata &    \nodata \\
        KP100920                  & 10.671 & 131.08151 & 20.80536 &  9.907  &  9.638  &  9.643  & J08441956+2048192\tablenotemark{b} &  0.4106  &   EB &    \nodata &    \nodata \\
        KP101034                  & 10.796 & 130.48279 & 19.68971 &  9.869  &  9.627  &  9.544  & J08415586+1941229         	     &  0.8894  & Puls &    \nodata &    \nodata \\
        KP101095                  & 10.871 & 133.01889 & 15.36116 & 10.206  & 10.011  &  9.977  & J08520453+1521401         	     &  0.4240  &   EB &    \nodata &    \nodata \\
        KP101194                  & 10.969 & 134.29383 & 17.97599 &  9.730  &  9.240  &  9.143  & J08571051+1758335         	     &  3.5604  & Puls &    \nodata &    \nodata \\
        KP101231                  & 11.000 & 134.29044 & 18.94559 &  9.794  &  9.391  &  9.272  & J08570970+1856441         	     &  0.2910  &   EB &    \nodata &    \nodata \\
        KP101257                  & 11.028 & 133.13490 & 16.80922 & 11.327  & 11.044  & 10.973  & J08523237+1648331         	     &  3.5740  &   EB &    \nodata &    \nodata \\
        KP101275\tablenotemark{a} & 11.053 & 130.47658 & 19.25742 & 10.026  &  9.732  &  9.643  & J08415437+1915267         	     &  0.8067  & Puls &    \nodata &    \nodata \\
        KP101456                  & 11.209 & 133.22672 & 21.98288 & 10.572  & 10.414  & 10.322  & J08525441+2158583         	     &  0.4351  & Puls &    \nodata &    \nodata \\
        KP101496                  & 11.253 & 133.20090 & 18.39749 &  9.254  &  8.428  &  8.211  & J08524821+1823509         	     &  \nodata &  LPV &    \nodata &    \nodata \\
	KP101511                  & 11.263 & 129.92462 & 14.08659 & 10.405  & 10.021  &  9.990  & J08394194+1405136         	     &  3.3393  &   EB &    \nodata &    \nodata \\
        KP101549                  & 11.289 & 132.56274 & 23.22313 & \nodata & \nodata & \nodata &           \nodata          	     &  \nodata &  LPV &    \nodata &    \nodata \\
        KP101674                  & 11.402 & 132.54651 & 23.61266 & \nodata & \nodata & \nodata &           \nodata          	     &  0.1612  & Puls &    \nodata &    \nodata \\
        KP101756                  & 11.464 & 130.16648 & 23.26186 & \nodata & \nodata & \nodata &           \nodata          	     &  0.2958  & Puls &     EF Cnc &      W UMa \\
        KP101831                  & 11.509 & 131.82932 & 20.83213 & 10.481  &  9.999  &  9.859  & J08471903+2049556\tablenotemark{b} &  \nodata &  LPV &    \nodata &    \nodata \\
        KP101896                  & 11.552 & 132.52862 & 23.52362 & \nodata & \nodata & \nodata &           \nodata          	     &  \nodata &  LPV &    \nodata &    \nodata \\
        KP101921                  & 11.573 & 132.90697 & 20.04515 & 10.261  &  9.730  &  9.642  & J08513767+2002425         	     &  0.5766  &   EB &    \nodata &    \nodata \\
        KP102135                  & 11.692 & 133.33712 & 22.75218 & 10.656  & 10.317  & 10.270  & J08532090+2245078         	     &  1.1039  & Puls &    \nodata &    \nodata \\
        KP102245                  & 11.749 & 132.76992 & 21.70140 & 10.895  & 10.622  & 10.585  & J08510478+2142050         	     &  0.5188  &   EB &    \nodata &    \nodata \\
        KP102395                  & 11.819 & 134.66732 & 17.24139 & 10.707  & 10.209  & 10.107  & J08584015+1714290         	     &  \nodata &  LPV &    \nodata &    \nodata \\
        KP102511                  & 11.878 & 131.20885 & 20.18795 &  9.815  &  9.174  &  8.992  & J08445012+2011166         	     &  0.5584  &   EB &    \nodata &    \nodata \\
        KP102540                  & 11.894 & 130.44375 & 14.87345 & 10.991  & 10.713  & 10.598  & J08414649+1452244         	     &  0.8213  & Puls &    \nodata &    \nodata \\
        KP102588                  & 11.916 & 132.71335 & 19.35729 & 11.746  & 11.554  & 11.495  & J08505120+1921262         	     &  1.3244  &   EB &  NSV 04269 &          V \\
        KP102705                  & 11.980 & 134.51968 & 22.63479 & 11.082  & 10.806  & 10.749  & J08580472+2238052         	     &  0.5978  & Puls &    \nodata &    \nodata \\
        KP102807                  & 12.030 & 132.66099 & 16.71456 & 10.993  & 10.610  & 10.500  & J08503863+1642524         	     &  0.8599  & Puls &    \nodata &    \nodata \\
        KP102811                  & 12.034 & 133.42089 & 22.34420 & 10.791  & 10.374  & 10.295  & J08534101+2220391         	     &  0.6102  &   EB &    \nodata &    \nodata \\
        KP102836                  & 12.047 & 134.25911 & 23.52545 & \nodata & \nodata & \nodata &           \nodata          	     &  3.1742  &   EB &    \nodata &    \nodata \\
        KP102908                  & 12.082 & 134.96979 & 17.47890 & 10.108  &  9.268  &  9.050  & J08595274+1728440         	     &  \nodata &  LPV &    \nodata &    \nodata \\
        KP102979                  & 12.112 & 130.63708 & 21.17889 & 11.707  & 11.658  & 11.643  & J08423289+2110440         	     &  0.8773  & Puls &    \nodata &    \nodata \\
        KP103010                  & 12.124 & 130.34124 & 18.13422 & 11.430  & 11.184  & 11.132  & J08412189+1808031         	     &  1.2578  &   EB &    \nodata &    \nodata \\
        KP103073\tablenotemark{a} & 12.150 & 130.02381 & 19.02520 & 10.658  & 10.149  & 10.009  & J08400571+1901307         	     &  0.8246  & Puls &    \nodata &    \nodata \\
        KP103143                  & 12.176 & 132.41259 & 23.80157 & \nodata & \nodata & \nodata &           \nodata          	     &  1.2542  &   EB &    \nodata &    \nodata \\
        KP103254                  & 12.226 & 133.24489 & 23.78463 & \nodata & \nodata & \nodata &           \nodata          	     &  0.5458  & Puls &     EZ Cnc &     RR Lyr \\
        KP103267                  & 12.229 & 133.18612 & 22.51454 & 10.059  &  9.452  &  9.219  & J08524466+2230523\tablenotemark{b} &  \nodata &  LPV &    \nodata &    \nodata \\
        KP103271                  & 12.232 & 131.87079 & 16.27255 & 11.116  & 10.659  & 10.573  & J08472898+1616211         	     &  0.8943  & Puls &    \nodata &    \nodata \\
        KP103285                  & 12.237 & 131.47646 & 19.58272 & 11.378  & 11.071  & 11.016  & J08455435+1934577         	     &  0.3550  &   EB &    \nodata &    \nodata \\
        KP103393                  & 12.280 & 132.90310 & 20.05444 & 11.801  & 11.626  & 11.603  & J08513674+2003159         	     &  0.5763  &   EB &    \nodata &    \nodata \\
        KP103585                  & 12.345 & 131.02744 & 18.51100 & 11.539  & 11.213  & 11.182  & J08440658+1830395         	     &  4.4082  &   EB &    \nodata &    \nodata \\
        KP103593                  & 12.348 & 134.67894 & 14.88607 & 11.406  & 11.075  & 10.993  & J08584294+1453098         	     &  0.3536  &   EB &    \nodata &    \nodata \\
        KP103608                  & 12.353 & 130.90608 & 19.93658 & 11.018  & 10.508  & 10.354  & J08433745+1956116         	     &  8.8821  & Puls &    \nodata &    \nodata \\
        KP103772                  & 12.410 & 130.51883 & 24.10952 & \nodata & \nodata & \nodata &           \nodata          	     &  0.2603  &   EB &    \nodata &    \nodata \\
        KP104115                  & 12.530 & 134.02845 & 14.64294 & 11.822  & 11.520  & 11.481  & J08560682+1438345         	     &  2.6984  &   EB &    \nodata &    \nodata \\
        KP104166                  & 12.549 & 130.52602 & 21.26102 & 10.554  &  9.810  &  9.626  & J08420624+2115396         	     &  \nodata &  LPV &    \nodata &    \nodata \\
        KP104174                  & 12.552 & 130.67742 & 21.41570 & 11.518  & 11.259  & 11.213  & J08424258+2124565         	     &  0.3637  &   EB &  NSV 04207 &          V \\
        KP104185                  & 12.557 & 132.05289 & 21.12052 & 11.272  & 10.940  & 10.843  & J08481269+2107138         	     &  0.2814  &   EB &     GW Cnc &      W UMa \\
        KP104317                  & 12.599 & 130.33974 & 19.00734 & 11.592  & 11.293  & 11.198  & J08412153+1900264         	     &  0.3464  &   EB &    \nodata &    \nodata \\
        KP104332                  & 12.602 & 135.08220 & 14.35794 & 11.736  & 11.271  & 11.162  & J09001972+1421285         	     &  0.2698  &   EB &    \nodata &    \nodata \\
        KP104424                  & 12.629 & 131.62549 & 22.65176 & 10.807  & 10.190  & 10.067  & J08463011+2239063         	     &  \nodata &  LPV &    \nodata &    \nodata \\
        KP104475                  & 12.644 & 132.03164 & 23.12175 & \nodata & \nodata & \nodata &           \nodata          	     &  2.4236  & Puls &    \nodata &    \nodata \\
        KP104572                  & 12.670 & 130.57458 & 16.39318 & 12.266  & 12.065  & 12.067  & J08421789+1623354         	     &  0.0865  & Puls &    \nodata &    \nodata \\
        KP104597                  & 12.676 & 130.52484 & 21.43683 & 11.868  & 11.697  & 11.646  & J08420596+2126125         	     &  0.4586  &   EB &    \nodata &    \nodata \\
        KP104630                  & 12.683 & 134.71800 & 21.07624 & 10.058  &  9.433  &  9.190  & J08585231+2104344         	     &  \nodata &  LPV &    \nodata &    \nodata \\
        KP104850                  & 12.741 & 132.90945 & 16.52710 & 10.424  &  9.556  &  9.346  & J08513826+1631375         	     &  \nodata &  LPV &    \nodata &    \nodata \\
        KP104855                  & 12.742 & 129.97762 & 19.82194 & 12.514  & 12.217  & 12.171  & J08395462+1949189         	     &  1.0929  &   EB &     RY Cnc &      EA/SD \\
        KP104917                  & 12.760 & 131.04394 & 22.25307 & 11.804  & 11.071  & 10.945  & J08441054+2215110\tablenotemark{b} &  0.7091  &   EB &    \nodata &    \nodata \\
        KP105076                  & 12.804 & 130.07613 & 16.60935 & 12.014  & 11.700  & 11.632  & J08401827+1636336         	     &  0.1489  & Puls &    \nodata &    \nodata \\
        KP105119                  & 12.815 & 130.94612 & 14.90738 & 12.433  & 12.299  & 12.305  & J08434706+1454265         	     &  0.7477  &   EB &    \nodata &    \nodata \\
        KP105369                  & 12.877 & 130.41766 & 14.80151 & 11.489  & 10.980  & 10.794  & J08414023+1448054         	     & 11.0424  & Puls &    \nodata &    \nodata \\
        KP105756                  & 12.971 & 135.26383 & 17.89894 & 11.273  & 11.051  & 10.922  & J09010331+1753561         	     &  \nodata &  LPV &    \nodata &    \nodata \\
        KP105772                  & 12.975 & 131.03903 & 17.06938 & 11.729  & 11.250  & 11.172  & J08440936+1704097         	     &  \nodata &  LPV &    \nodata &    \nodata \\
        KP105793                  & 12.979 & 132.74935 & 13.96247 & 12.230  & 11.954  & 11.909  & J08505984+1357448         	     &  0.3216  &   EB &    \nodata &    \nodata \\
        KP105899                  & 13.005 & 133.45578 & 20.85212 & 11.412  & 10.845  & 10.747  & J08534938+2051076         	     &  0.8162  & Puls &    \nodata &    \nodata \\
        KP106106                  & 13.054 & 135.21007 & 23.62369 & \nodata & \nodata & \nodata &           \nodata          	     &  0.6997  &   EB &    \nodata &    \nodata \\
        KP106218                  & 13.077 & 133.14519 & 22.48431 & 11.156  & 10.453  & 10.309  & J08523484+2229035         	     &  \nodata &  LPV &    \nodata &    \nodata \\
        KP106227                  & 13.080 & 133.37819 & 14.13207 & 11.563  & 11.003  & 10.839  & J08533076+1407554         	     &  \nodata &  LPV &    \nodata &    \nodata \\
        KP106319                  & 13.096 & 130.44779 & 21.97142 & 12.161  & 11.909  & 11.847  & J08414746+2158171         	     &  0.1846  & Puls &    \nodata &    \nodata \\
        KP106351                  & 13.102 & 134.54297 & 15.80525 & 12.112  & 11.941  & 11.944  & J08581031+1548188         	     &  0.5430  & Puls &     AN Cnc &     RR Lyr \\
        KP106452                  & 13.122 & 131.34256 & 15.27477 & 12.187  & 11.994  & 11.951  & J08452221+1516291         	     &  0.5246  & Puls &     CQ Cnc &     RR Lyr \\
        KP106608                  & 13.157 & 133.67184 & 19.11521 & 12.352  & 12.038  & 11.998  & J08544124+1906547\tablenotemark{b} &  0.3396  &   EB &    \nodata &    \nodata \\
        KP106885                  & 13.213 & 133.82978 & 16.43977 & 11.455  & 10.694  & 10.515  & J08551914+1626231         	     &  \nodata &  LPV &    \nodata &    \nodata \\
        KP107014                  & 13.234 & 130.22716 & 14.42231 & 12.345  & 12.011  & 12.019  & J08405451+1425203         	     &  0.2465  & Puls &    \nodata &    \nodata \\
        KP107531                  & 13.333 & 129.98160 & 23.33538 & \nodata & \nodata & \nodata &           \nodata          	     &  0.5516  &   EB &    \nodata &    \nodata \\
        KP107924                  & 13.408 & 135.19565 & 15.42220 & 12.941  & 12.796  & 12.724  & J09004695+1525199         	     &  0.5268  &   EB &    \nodata &    \nodata \\
        KP108285                  & 13.473 & 130.92216 & 18.42757 & 12.735  & 12.477  & 12.425  & J08434131+1825392         	     &  0.3360  &   EB &    \nodata &    \nodata \\
        KP108858                  & 13.569 & 130.25118 & 13.67656 & 12.607  & 12.213  & 12.119  & J08410028+1340356         	     &  0.2986  &   EB &    \nodata &    \nodata \\
        KP109198                  & 13.619 & 133.45391 & 21.49120 & 13.066  & 12.953  & 12.946  & J08534893+2129283         	     &  1.7433  &   EB &    \nodata &    \nodata \\
        KP109247                  & 13.625 & 133.88598 & 14.54473 & 13.160  & 12.925  & 12.885  & J08553263+1432410         	     &  0.4205  &   EB &    \nodata &    \nodata \\
        KP110021                  & 13.734 & 132.80862 & 16.07745 & 13.036  & 12.801  & 12.732  & J08511406+1604388         	     &  0.3860  &   EB &    \nodata &    \nodata \\
        KP110124                  & 13.751 & 134.72799 & 15.36940 & 13.450  & 13.295  & 13.300  & J08585471+1522098         	     &  0.0576  & Puls &    \nodata &    \nodata \\
        KP110177                  & 13.760 & 130.17535 & 14.98393 & 12.410  & 11.867  & 11.713  & J08404208+1459021         	     &  0.4420  &   EB &    \nodata &    \nodata \\
        KP110305                  & 13.777 & 130.52273 & 22.41861 & 12.483  & 12.007  & 11.900  & J08420545+2225069         	     &  0.2719  &   EB &    \nodata &    \nodata \\
        KP110392                  & 13.792 & 130.19981 & 15.41455 & 13.402  & 13.115  & 13.061  & J08404795+1524523         	     &  0.6004  & Puls &    \nodata &    \nodata \\
        KP110617                  & 13.826 & 131.51008 & 15.44463 & \nodata & \nodata & \nodata &           \nodata          	     &  0.3395  &   EB &    \nodata &    \nodata \\
        KP110674                  & 13.837 & 133.33587 & 23.52222 & \nodata & \nodata & \nodata &           \nodata          	     & 11.0384  &   EB &    \nodata &    \nodata \\
        KP110871                  & 13.864 & 130.70418 & 15.39379 & 12.524  & 12.010  & 11.957  & J08424900+1523376         	     &  0.2836  &   EB &    \nodata &    \nodata \\
        KP110876                  & 13.865 & 131.03416 & 16.84885 & 12.866  & 12.492  & 12.420  & J08440819+1650558         	     &  0.3013  &   EB &    \nodata &    \nodata \\
        KP111363                  & 13.933 & 134.00672 & 18.27980 & \nodata & \nodata & \nodata &           \nodata          	     &  \nodata &  LPV &    \nodata &    \nodata \\
        KP112203                  & 14.042 & 132.87884 & 21.75790 & 13.009  & 12.588  & 12.521  & J08513092+2145284         	     &  0.3567  &   EB &    \nodata &    \nodata \\
        KP112538                  & 14.086 & 129.95615 & 15.42027 & 13.304  & 12.770  & 12.688  & J08394947+1525129         	     &  0.2722  &   EB &    \nodata &    \nodata \\
        KP113453                  & 14.197 & 132.70565 & 16.31045 & 13.683  & 13.550  & 13.455  & J08504935+1618376         	     &  0.3395  & Puls &    \nodata &    \nodata \\
        KP113808                  & 14.236 & 131.51525 & 20.96654 & 13.211  & 12.920  & 12.805  & J08460366+2057595         	     &  0.3182  &   EB &    \nodata &    \nodata \\
        KP114383                  & 14.301 & 131.15271 & 21.75311 & 13.371  & 13.095  & 13.007  & J08443665+2145111         	     &  0.3478  &   EB &    \nodata &    \nodata \\
        KP114757                  & 14.344 & 133.70368 & 20.10852 & 13.602  & 12.676  & 11.830  & J08544888+2006306         	     &  \nodata &  LPV &    \nodata &    \nodata \\
        KP115386                  & 14.408 & 134.78593 & 23.89257 & \nodata & \nodata & \nodata &           \nodata          	     &  0.2981  & Puls &    \nodata &    \nodata \\
        KP115639                  & 14.434 & 131.93564 & 23.76530 & \nodata & \nodata & \nodata &           \nodata          	     &  0.5692  &   EB &    \nodata &    \nodata \\
        KP115973                  & 14.469 & 135.41519 & 16.40212 & 13.531  & 13.232  & 13.135  & J09013964+1624076         	     &  \nodata &  LPV &    \nodata &    \nodata \\
        KP118312                  & 14.698 & 132.30275 & 20.90523 & 13.923  & 13.617  & 13.445  & J08491266+2054188         	     &  \nodata &  LPV &    \nodata &    \nodata \\
        KP118899                  & 14.754 & 134.88939 & 21.16955 & 13.444  & 12.955  & 12.812  & J08593345+2110103         	     &  0.2897  &   EB &    \nodata &    \nodata \\
        KP119499                  & 14.808 & 135.31678 & 19.63988 & 13.468  & 12.922  & 12.869  & J09011602+1938235         	     &  \nodata &  LPV &    \nodata &    \nodata \\
        KP121436                  & 14.975 & 130.64868 & 20.64923 & 13.881  & 13.436  & 13.403  & J08423568+2038572         	     &  \nodata &  LPV &    \nodata &    \nodata \\
        KP126266                  & 15.351 & 132.71161 & 19.68410 & 14.129  & 13.679  & 13.526  & J08505078+1941027         	     &  \nodata &  LPV &    \nodata &    \nodata \\
        KP127604                  & 15.455 & 131.49181 & 19.93790 & 14.733  & 14.361  & 14.183  & J08455803+1956164         	     &  \nodata &  LPV &    \nodata &    \nodata \\
        KP127745                  & 15.466 & 132.14931 & 20.92373 & 14.871  & 14.607  & 14.455  & J08483583+2055254         	     &  \nodata &  LPV &    \nodata &    \nodata \\
        KP128381                  & 15.511 & 132.94836 & 16.20497 & 14.488  & 14.005  & 13.915  & J08514760+1612178         	     &  \nodata &  LPV &    \nodata &    \nodata \\
        KP130733                  & 15.689 & 131.45529 & 20.12358 & 14.260  & 13.725  & 13.654  & J08454926+2007248         	     &  \nodata &  LPV &    \nodata &    \nodata \\
        KP131942                  & 15.791 & 135.36885 & 18.02716 & 14.784  & 14.285  & 14.315  & J09012852+1801377         	     &  \nodata &  LPV &    \nodata &    \nodata \\
        KP133934                  & 16.024 & 131.81019 & 19.93585 & 14.405  & 13.792  & 13.684  & J08471444+1956090         	     &  \nodata &  LPV &    \nodata &    \nodata \\
        KP200074                  &  9.079 & 129.40268 & 16.02737 &  5.746  &  4.835  &  4.674  & J08373664+1601385         	     &  \nodata &  LPV &    \nodata &    \nodata \\
        KP200152                  &  9.711 & 128.95538 & 21.21765 &  7.873  &  7.189  &  7.010  & J08354929+2113035         	     &  \nodata &  LPV &    \nodata &    \nodata \\
        KP200188                  & 10.019 & 129.38447 & 23.55798 & \nodata & \nodata & \nodata &           \nodata          	     & 10.0591  &   EB &     RU Cnc &   EA/DS/RS \\
        KP200191                  & 10.024 & 128.87199 & 15.14900 &  7.086  &  6.216  &  5.903  & J08352927+1508563         	     &  \nodata &  LPV &    \nodata &    \nodata \\
        KP200210                  & 10.116 & 129.65732 & 22.76973 &  7.908  &  7.128  &  6.893  & J08383775+2246110         	     &  \nodata &  LPV &    \nodata &    \nodata \\
        KP200224                  & 10.184 & 128.88365 & 17.05935 &  8.954  &  8.538  &  8.435  & J08353207+1703336         	     & 12.1114  & Puls &    \nodata &    \nodata \\
        KP200244                  & 10.312 & 128.68203 & 23.04457 & \nodata & \nodata & \nodata &           \nodata          	     &  \nodata &  LPV &    \nodata &    \nodata \\
        KP200257                  & 10.384 & 129.60230 & 20.50270 &  8.466  &  7.701  &  7.542  & J08382455+2030097         	     &  \nodata &  LPV &    \nodata &    \nodata \\
        KP200262                  & 10.408 & 129.66094 & 21.42135 &  8.969  &  8.400  &  8.233  & J08383862+2125168         	     &  \nodata &  LPV &    \nodata &    \nodata \\
        KP200270                  & 10.453 & 128.69299 & 17.76642 &  9.779  &  9.611  &  9.576  & J08344631+1745591         	     &  0.7277  &   EB &    \nodata &    \nodata \\
        KP200312                  & 10.600 & 129.46961 & 16.36575 &  9.745  &  9.498  &  9.449  & J08375270+1621566         	     &  2.1302  &   EB &    \nodata &    \nodata \\
        KP200490                  & 11.146 & 128.55033 & 16.73073 & 10.514  & 10.351  & 10.347  & J08341207+1643506         	     &  0.1297  & Puls &    \nodata &    \nodata \\
        KP200596                  & 11.385 & 128.73938 & 19.91683 & 10.245  &  9.884  &  9.801  & J08345745+1955005         	     &  0.3234  &   EB &    \nodata &    \nodata \\
        KP200643                  & 11.489 & 128.56992 & 13.98220 & 10.487  & 10.231  & 10.176  & J08341678+1358559         	     &  0.3866  &   EB &    \nodata &    \nodata \\
        KP200711                  & 11.642 & 129.94680 & 14.29008 & 11.065  & 10.910  & 10.862  & J08394723+1417242         	     &  0.2609  & Puls &    \nodata &    \nodata \\
        KP200738                  & 11.701 & 129.30049 & 13.84782 & 11.087  & 10.834  & 10.804  & J08371211+1350521         	     &  0.4046  &   EB &    \nodata &    \nodata \\
        KP200757\tablenotemark{a} & 11.745 & 129.90634 & 18.17039 & 10.763  & 10.321  & 10.237  & J08393752+1810134         	     &  1.1150  & Puls &    \nodata &    \nodata \\
        KP201187                  & 12.361 & 128.96040 & 20.55906 & 11.273  & 10.925  & 10.816  & J08355049+2033326         	     &  1.4783  &   EB &    \nodata &    \nodata \\
        KP201350                  & 12.543 & 129.40514 & 14.59862 & 11.346  & 10.913  & 10.823  & J08373723+1435550         	     &  0.8728  &   EB &    \nodata &    \nodata \\
        KP201399                  & 12.596 & 129.91210 & 13.72237 & 11.930  & 11.726  & 11.641  & J08393890+1343205         	     &  0.2163  & Puls &    \nodata &    \nodata \\
        KP201426                  & 12.636 & 128.77766 & 15.86104 & 11.954  & 11.627  & 11.569  & J08350663+1551397         	     &  1.5944  &   EB &    \nodata &    \nodata \\
        KP201594                  & 12.800 & 129.32782 & 17.04804 & 11.297  & 10.709  & 10.598  & J08371867+1702529         	     &  \nodata &  LPV &    \nodata &    \nodata \\
        KP201632                  & 12.837 & 129.79576 & 16.79492 & 12.113  & 11.892  & 11.800  & J08391098+1647417         	     &  0.3361  &   EB &    \nodata &    \nodata \\
        KP202047                  & 13.178 & 128.80993 & 23.68985 & \nodata & \nodata & \nodata &           \nodata          	     &  0.6916  &   EB &    \nodata &    \nodata \\
        KP202125                  & 13.224 & 129.87202 & 23.58952 & \nodata & \nodata & \nodata &           \nodata          	     &  0.3682  &   EB &    \nodata &    \nodata \\
        KP202179                  & 13.269 & 129.64794 & 13.70668 & 12.749  & 12.599  & 12.559  & J08383550+1342240         	     &  0.6204  &   EB &    \nodata &    \nodata \\
        KP202440                  & 13.435 & 128.73262 & 22.55996 & 12.599  & 12.355  & 12.340  & J08345582+2233358         	     &  2.7325  &   EB &    \nodata &    \nodata \\
        KP202483                  & 13.459 & 129.83068 & 15.51133 & 12.599  & 12.285  & 12.237  & J08391936+1530407         	     &  \nodata &  LPV &    \nodata &    \nodata \\
        KP202655                  & 13.556 & 129.50897 & 16.99027 & 12.423  & 12.060  & 11.952  & J08380215+1659249         	     &  0.3784  &   EB &  NSV 04158 &          V \\
        KP202778                  & 13.608 & 128.97139 & 20.91898 & 12.825  & 12.476  & 12.425  & J08355313+2055083         	     &  1.3931  &   EB &    \nodata &    \nodata \\
        KP202994                  & 13.717 & 129.72676 & 21.17214 & 13.013  & 12.676  & 12.629  & J08385442+2110197         	     &  \nodata &  LPV &    \nodata &    \nodata \\
        KP203288                  & 13.858 & 129.75875 & 15.25845 & 13.352  & 13.208  & 13.203  & J08390210+1515304         	     &  0.5163  &   EB &    \nodata &    \nodata \\
        KP204370                  & 14.282 & 128.39201 & 22.06084 & 13.515  & 13.231  & 13.187  & J08333408+2203390         	     &  0.1534  & Puls &    \nodata &    \nodata \\
        KP204605                  & 14.381 & 128.40400 & 18.16221 & 13.367  & 13.008  & 13.013  & J08333696+1809439         	     &  0.1667  & Puls &    \nodata &    \nodata \\
        KP206403                  & 14.927 & 129.55466 & 17.42242 & 13.564  & 13.086  & 13.007  & J08381311+1725207         	     &  0.2525  &   EB &    \nodata &    \nodata \\
        KP300063                  &  9.219 & 128.25422 & 17.22665 &  7.909  &  7.416  &  7.310  & J08330101+1713359         	     &  \nodata &  LPV &    \nodata &    \nodata \\
        KP300133                  &  9.785 & 127.86619 & 19.88428 &  8.335  &  7.739  &  7.619  & J08312788+1953034         	     &  0.1089  & Puls &    \nodata &    \nodata \\
        KP300135                  &  9.788 & 128.12730 & 15.82408 &  8.060  &  7.456  &  7.307  & J08323055+1549266         	     &  0.8270  & Puls &     FR Cnc &     BY Dra \\
        KP300161                  &  9.982 & 128.00962 & 19.60058 &  9.263  &  8.970  &  8.895  & J08320230+1936020         	     &  1.4661  &   EB &    \nodata &    \nodata \\
        KP300277                  & 10.581 & 127.41380 & 17.28352 &  9.384  &  8.939  &  8.853  & J08293931+1717006         	     &  1.3236  &   EB &     FF Cnc &      Algol \\
        KP300434                  & 11.088 & 127.15242 & 21.93378 &  7.999  &  7.135  &  6.797  & J08283658+2156016         	     &  \nodata &  LPV &    \nodata &    \nodata \\
        KP300526                  & 11.290 & 127.00145 & 22.93127 & \nodata & \nodata & \nodata &           \nodata          	     &  0.3154  &   EB &    \nodata &    \nodata \\
        KP300603                  & 11.482 & 126.91865 & 19.26223 &  7.825  &  7.091  &  6.727  & J08274047+1915440         	     &  \nodata &  LPV &     GV Cnc &      SR SR \\
        KP300608                  & 11.496 & 127.35241 & 14.36576 &  9.887  &  9.313  &  9.184  & J08292457+1421567         	     &  7.6757  & Puls &    \nodata &    \nodata \\
        KP300656                  & 11.603 & 127.85822 & 20.91667 & 10.970  & 10.680  & 10.582  & J08312597+2055000         	     &  2.2290  &   EB &    \nodata &    \nodata \\
        KP300678                  & 11.658 & 127.31299 & 18.38538 &  8.748  &  7.709  &  7.067  & J08291511+1823073         	     &  \nodata &  LPV &    \nodata &    \nodata \\
        KP300976                  & 12.169 & 127.09836 & 19.97421 &  9.123  &  8.204  &  7.873  & J08282360+1958271         	     &  \nodata &  LPV &    \nodata &    \nodata \\
        KP301090                  & 12.320 & 127.51882 & 16.73953 & 11.129  & 10.777  & 10.691  & J08300451+1644223         	     &  0.8230  & Puls &    \nodata &    \nodata \\
        KP301148                  & 12.393 & 128.17451 & 23.55348 & \nodata & \nodata & \nodata &           \nodata          	     &  0.3113  &   EB &    \nodata &    \nodata \\
        KP301208                  & 12.453 & 127.73636 & 21.16272 & 11.874  & 11.676  & 11.621  & J08305672+2109457         	     &  0.2175  & Puls &    \nodata &    \nodata \\
        KP301312                  & 12.545 & 127.28293 & 20.57023 & 11.742  & 11.484  & 11.428  & J08290790+2034128         	     &  0.8214  &   EB &    \nodata &    \nodata \\
        KP301510                  & 12.736 & 127.20627 & 18.42707 & 12.006  & 11.695  & 11.621  & J08284950+1825374         	     &  0.1840  & Puls &    \nodata &    \nodata \\
        KP301583                  & 12.805 & 127.09350 & 19.16872 & 11.616  & 11.278  & 11.138  & J08282244+1910073         	     &  0.3393  &   EB &    \nodata &    \nodata \\
        KP301835                  & 13.020 & 127.01759 & 17.99201 & 11.593  & 11.050  & 10.914  & J08280422+1759312         	     &  0.2393  &   EB &    \nodata &    \nodata \\
        KP301980                  & 13.128 & 127.99386 & 17.38058 & 11.722  & 11.175  & 11.067  & J08315852+1722500         	     &  0.1390  & Puls &    \nodata &    \nodata \\
        KP302158                  & 13.233 & 127.90973 & 15.20447 & 12.264  & 11.932  & 11.841  & J08313833+1512160         	     &  0.2626  &   EB &    \nodata &    \nodata \\
        KP302288                  & 13.300 & 127.25518 & 20.34976 & 12.714  & 12.432  & 12.359  & J08290124+2020591         	     &  0.5335  &   EB &    \nodata &    \nodata \\
        KP302426                  & 13.382 & 128.43628 & 15.32971 & 13.087  & 12.565  & 12.511  & J08334470+1519469\tablenotemark{b} &  0.2975  &   EB &    \nodata &    \nodata \\
        KP302468                  & 13.405 & 128.16966 & 17.84913 & 12.984  & 12.788  & 12.753  & J08324071+1750568         	     &  0.2249  & Puls &    \nodata &    \nodata \\
        KP302519                  & 13.437 & 127.99812 & 21.41259 & 12.441  & 12.117  & 12.040  & J08315954+2124453         	     &  2.2423  &   EB &    \nodata &    \nodata \\
        KP302538                  & 13.448 & 128.07795 & 17.83218 & 12.612  & 12.286  & 12.239  & J08321870+1749558         	     &  0.3481  &   EB &    \nodata &    \nodata \\
        KP303144                  & 13.747 & 128.28988 & 23.46403 & \nodata & \nodata & \nodata &           \nodata          	     &  0.4900  & Puls &    \nodata &    \nodata \\
        KP303492                  & 13.890 & 127.32312 & 21.34491 & 13.578  & 13.279  & 13.272  & J08291754+2120416         	     &  \nodata &  LPV &    \nodata &    \nodata \\
        KP309546                  & 15.612 & 128.09182 & 18.63418 & 14.636  & 14.039  & 13.976  & J08322203+1838030         	     &  0.1089  & Puls &    \nodata &    \nodata \\
        KP400130                  &  9.412 & 124.81942 & 14.45165 &  6.620  &  5.790  &  5.493  & J08191666+1427059         	     &  \nodata &  LPV &    \nodata &    \nodata \\
        KP400137                  &  9.437 & 126.15084 & 14.65195 &  6.770  &  5.899  &  5.640  & J08243620+1439070         	     &  \nodata &  LPV &    \nodata &    \nodata \\
        KP400480                  & 10.765 & 125.68731 & 16.52782 &  7.883  &  7.042  &  6.744  & J08224495+1631401         	     &  \nodata &  LPV &    \nodata &    \nodata \\
        KP400729                  & 11.230 & 125.98279 & 21.18908 &  9.501  &  8.937  &  8.728  & J08235586+2111206         	     &  7.8954  & Puls &    \nodata &    \nodata \\
        KP400757                  & 11.269 & 126.49398 & 20.24876 & 10.448  & 10.223  & 10.163  & J08255855+2014555         	     &  0.9744  &   EB &    \nodata &    \nodata \\
        KP400793                  & 11.325 & 125.67915 & 19.44957 & 10.360  & 10.051  & 10.002  & J08224299+1926584        	     &  0.2800  &   EB &    \nodata &    \nodata \\
        KP400838                  & 11.399 & 125.71576 & 16.03616 & 10.622  & 10.351  & 10.301  & J08225178+1602101         	     &  0.8538  &   EB &    \nodata &    \nodata \\
        KP400943                  & 11.553 & 126.72200 & 21.46277 & 10.248  &  9.767  &  9.652  & J08265327+2127459         	     &  0.6944  & Puls &    \nodata &    \nodata \\
        KP400955                  & 11.566 & 124.85908 & 17.83969 & 10.517  & 10.228  & 10.126  & J08192617+1750228         	     &  0.7844  &   EB &    \nodata &    \nodata \\
        KP401088                  & 11.721 & 125.60951 & 20.98303 & 10.742  & 10.545  & 10.459  & J08222628+2058589         	     &  0.5609  &   EB &    \nodata &    \nodata \\
        KP401140                  & 11.791 & 126.57688 & 23.25369 & 11.348  & 11.184  & 11.136  & J08261845+2315132         	     &  0.7178  &   EB &  NSV 04069 &          V \\
        KP401180                  & 11.836 & 125.68698 & 19.46563 & 10.970  & 10.771  & 10.705  & J08224487+1927562         	     &  0.3257  &   EB &    \nodata &    \nodata \\
        KP401189                  & 11.845 & 126.57647 & 20.88048 & 11.134  & 10.925  & 10.859  & J08261835+2052497         	     &  0.4180  &   EB &     EH Cnc &      W UMa \\
        KP401228                  & 11.884 & 126.16900 & 18.70081 & 10.961  & 10.749  & 10.635  & J08244055+1842029         	     &  1.7831  &   EB &    \nodata &    \nodata \\
        KP401417                  & 12.065 & 124.71795 & 17.72634 & \nodata & \nodata & \nodata &           \nodata          	     &  0.7835  &   EB &    \nodata &    \nodata \\
        KP401692                  & 12.282 & 124.87608 & 17.56528 & \nodata & \nodata & \nodata &           \nodata          	     &  1.2360  &   EB &    \nodata &    \nodata \\
        KP401909                  & 12.450 & 125.93575 & 14.33630 & 11.870  & 11.659  & 11.599  & J08234458+1420106         	     &  0.3358  &   EB &    \nodata &    \nodata \\
        KP401910                  & 12.451 & 126.80627 & 17.67660 & 11.262  & 10.853  & 10.726  & J08271350+1740357         	     &  3.2616  &   EB &    \nodata &    \nodata \\
        KP402009                  & 12.510 & 126.62874 & 17.04807 & 11.173  & 10.712  & 10.598  & J08263089+1702530         	     &  0.2671  &   EB &    \nodata &    \nodata \\
        KP402737                  & 12.924 & 126.75819 & 18.81549 & 12.471  & 12.352  & 12.291  & J08270196+1848557         	     &  0.8880  &   EB &    \nodata &    \nodata \\
        KP402860                  & 12.978 & 125.85141 & 23.48793 & 12.271  & 12.009  & 11.960  & J08232433+2329165         	     &  0.5166  &   EB &    \nodata &    \nodata \\
        KP402888                  & 12.990 & 126.61546 & 23.15503 & \nodata & \nodata & \nodata &           \nodata          	     &  0.9017  &   EB &    \nodata &    \nodata \\
        KP404664                  & 13.668 & 126.70888 & 16.38223 & 10.496  &  9.720  &  9.518  & J08265013+1622560         	     &  \nodata &  LPV &    \nodata &    \nodata \\
        KP404675                  & 13.670 & 125.72269 & 20.52251 & 12.953  & 12.608  & 12.538  & J08225344+2031210         	     &  0.3067  &   EB &    \nodata &    \nodata \\
        KP404726                  & 13.687 & 125.37290 & 20.86872 & 12.796  & 12.579  & 12.566  & J08212949+2052073         	     &  0.3759  &   EB &    \nodata &    \nodata \\
        KP405535                  & 13.923 & 126.53358 & 18.81866 & 13.382  & 13.041  & 12.940  & J08260805+1849071         	     &  \nodata &  LPV &    \nodata &    \nodata \\
        KP407282                  & 14.342 & 125.03220 & 21.74567 & 13.035  & 12.618  & 12.516  & J08200772+2144444         	     &  0.3051  &   EB &    \nodata &    \nodata \\
        KP413347                  & 15.417 & 126.70529 & 20.46176 & 14.494  & 14.166  & 14.157  & J08264926+2027423         	     &  \nodata &  LPV &    \nodata &    \nodata \\
        KP414644                  & 15.648 & 125.77979 & 17.68299 & 14.464  & 14.020  & 13.919  & J08230714+1740587         	     &  \nodata &  LPV &    \nodata &    \nodata \\
\hline									    
\enddata								    
\tablenotetext{a}{Members of the Praesepe cluster.}			    
\tablenotetext{b}{Stars matched to more than one 2MASS source 
within the 9\farcs5 matching radius.  The 2MASS ID and $JHK$ colors in the table are for the closest 
match within the radius.}
\end{deluxetable}
        \clearpage
        \end{landscape}

\begin{deluxetable}{rrrrrcr}
\tablecaption{Properties of KELT Transit Candidates  \label{tab:tprop}}
\tablewidth{0pc}
\tablehead{
\colhead{KELT ID}            &
\colhead{2MASS ID}  &
\colhead{RA }  &
\colhead{Dec }       &
\colhead{$R_K$ }       &
\colhead{$J-K$}     &
\colhead{Period} \\
\colhead{}   &
\colhead{}   &
\colhead{(J2000.0)}    & 
\colhead{(J2000.0)}    & 
\colhead{mag}      &
\colhead{(2MASS)}    &
\colhead{(days)}    
}
\startdata
KP102662 & J08525435+1447557 & 133.22647 & 14.79883 & 11.96 & 0.31 & 1.8558 \\
KP102791 & J08465697+1603190 & 131.73742 & 16.05529 & 12.02 & 0.30 & 3.0227 \\
KP200924 & J08362287+2045100 & 129.09531 & 20.75279 & 12.05 & 0.57 & 0.6246 \\
KP103126 & J08530235+2045045 & 133.25982 & 20.75127 & 12.17 & 0.38 & 0.4984 \\
\enddata
\end{deluxetable}

\begin{deluxetable}{rrr}
\tablecaption{Radial Velocities for KP102791  \label{tab:radvel}}
\tablewidth{0pc}
\tablehead{
\colhead{HJD}            &
\colhead{$V_{\rm rad}$}  &
\colhead{$\sigma(V_{\rm rad})$}  \\
\colhead{(days)}       &
\colhead{(\kms)}       &
\colhead{(\kms)}}
\startdata
24454108.8575 &$  34.95 $&  1.91 \\
24454127.7863 &$  37.87 $&  1.77 \\
24454135.8422 &$   8.25 $&  1.83 \\
24454137.8649 &$ -70.69 $&  1.84 \\
24454162.7826 &$ -27.46 $&  1.92 \\
24454165.7497 &$ -28.09 $&  1.82 \\
24454166.7061 &$  58.08 $&  1.36 \\
24454190.7266 &$  57.66 $&  1.41 \\
24454191.6968 &$ -19.43 $&  1.89 \\
24454192.6877 &$ -58.26 $&  2.43 \\
24454193.7257 &$  52.89 $&  1.64 \\
24454194.6724 &$ -11.43 $&  1.82 \\
24454195.7780 &$ -50.40 $&  1.71 \\
24454196.6919 &$  49.67 $&  2.12 \\
\enddata
\end{deluxetable}

\end{document}